\documentclass[12pt,preprint]{aastex}

\shorttitle{Horizontal Branch Morphology}
\shortauthors{Dotter et a.}

\slugcomment{To appear in The Astrophyical Journal}

\newcommand{\rgc}{\mathrm{R_{GC}}}
\newcommand{\rh} {\mathrm{R_h}}
\newcommand{\rt} {\mathrm{R_t}}
\newcommand{\dvi}{\Delta(V-I)}
\newcommand{\mv} {M_V}
\newcommand{\feh}{\mathrm{[Fe/H]}}
\newcommand{\afe}{\mathrm{[\alpha/Fe]}}
\newcommand{\thb}{\mathrm{T_{eff}(HB)}}
\newcommand{\dfm}{\mathrm{\Delta(Fit-Measured)}}
\begin{document}

\title{ The ACS Survey of Galactic Globular Clusters. IX.\thanks{
Based on observations with the NASA/ESA {\it Hubble Space Telescope},
obtained at the Space Telescope Science Institute, which is operated
by AURA, Inc., under NASA contract NAS 5-26555, under program GO-10775.
}\\ Horizontal Branch Morphology and the Second Parameter Phenomenon}

\author{Aaron Dotter} \affil{Department of Physics and Astronomy, University of Victoria, PO Box 3055, STN CSC, Victoria, BC, V8W 3P6 Canada\email{dotter@uvic.ca}}
\author{Ata Sarajedini} \affil{Department of Astronomy, University of Florida, 211 Bryant Space Science Center, Gainesville, FL 32611 \email{ata@astro.ufl.edu}}
\author{Jay Anderson} \affil{Space Telescope Science Institute, 3700 San Martin Drive, Baltimore, MD 21218, USA}
\author{Antonio Aparicio} \affil{Instituto de Astrof\'isica de Canarias, V\'ia L\'actea s/n, E-38200 La Laguna, Spain}
\author{Luigi R. Bedin} \affil{Space Telescope Science Institute, 3700 San Martin Drive, Baltimore, MD 21218, USA}
\author{Brian Chaboyer} \affil{Department of Physics and Astronomy, Dartmouth College, 6127 Wilder Laboratory, Hanover, NH 03755}
\author{Steven Majewski} \affil{Dept. of Astronomy, University of Virginia, P.O. Box 400325, Charlottesville, VA 22904-4325}
\author{A. Mar\'in-Franch} \affil{Instituto de Astrof\'isica de Canarias, V\'ia L\'actea s/n, E-38200 La Laguna, Spain }
\author{Antonino Milone} \affil{Dipartimento di Astronomia, Universit\`a di Padova, 35122 Padova, Italy }
\author{Nathaniel Paust} \affil{Space Telescope Science Institute, 3700 San Martin Drive, Baltimore MD 21218}
\author{Giampaolo Piotto} \affil{Dipartimento di Astronomia, Universit\`a di Padova, 35122 Padova, Italy }
\author{I. Neill Reid} \affil{Space Telescope Science Institute, 3700 San Martin Drive, Baltimore MD 21218}
\author{Alfred Rosenberg} \affil{Instituto de Astrof\'isica de Canarias, V\'ia L\'actea s/n, E-38200 La Laguna, Spain}
\author{Michael Siegel} \affil{Department of Astronomy and Astrophysics, Pennsylvania State University, 525 Davey Laboratory, State College, PA  16801}

\begin{abstract}
The horizontal branch (HB) morphology of globular clusters 
(GCs) is most strongly influenced by metallicity. The second parameter phenomenon, first 
described in the 1960's, acknowledges that metallicity alone is not enough to describe 
the HB morphology of all GCs. In particular, astronomers noticed that the outer Galactic 
halo contains GCs with redder HBs at a given metallicity than are found inside the Solar 
circle. Thus, at least a second parameter was required to characterize HB morphology. 
While the term `second parameter' has since come to be used in a broader context, its 
identity with respect to the original problem has not been conclusively determined. Here 
we analyze the median color difference between the HB and the red giant branch (RGB), 
hereafter denoted $\dvi$, measured from Hubble Space Telescope (HST) Advanced Camera for 
Surveys (ACS) photometry of 60 GCs within $\sim$20 kpc of the Galactic Center. 
Analysis of this homogeneous data set reveals that, after the influence of metallicity 
has been removed from the data, the correlation between $\dvi$ and age is stronger than
that of any other parameter considered. Expanding the sample to include HST ACS and 
Wide Field Planetary Camera 2 (WFPC2) photometry of the 6 most distant Galactic GCs
lends additional support to the correlation between $\dvi$ and age. This result is robust 
with respect to the adopted metallicity scale and the method of age determination, but must 
bear the caveat that high quality, detailed abundance information is not available for a 
significant fraction of the sample. Furthermore, when a subset of GCs with similar 
metallicities and ages are considered, a correlation between $\dvi$ and central luminosity 
density is exposed. With respect to the existence of GCs with anomalously red HBs at a 
given metallicity, we conclude that age is the second parameter and central density is 
most likely the third. Important problems related to HB morphology in GCs, notably 
multi-modal distributions and faint blue tails, remain to be explained.
\end{abstract}


\section{Introduction\label{intro}}

It has been clear for decades that metallicity is the most influential factor governing 
the HB morphologies of Galactic GCs: metallicity is the first parameter.  The earliest 
photographic color-magnitude diagrams (CMDs) of GCs by, e.g.\,\citet{ar52,sa53}, revealed
that HB stars in metal-rich GCs tend to lie on the red side of the RR Lyrae instability 
strip while HB stars in metal-poor GCs lie primarily on the blue side. As more and
more CMDs were assembled, exceptions to this rule were uncovered. \citet{sa60} noted 
that M~13 and M~22 display HB morphologies appropriate for metal-poor GCs, despite the 
fact both GCs were of intermediate metallicity, and suggested that a difference in age might be responsible. 
\citet{sa67} noticed that the HB of NGC~7006 is redder than either M~13 or M~3, despite the 
fact that all three GCs appeared to have very similar metallicities. Such anomalies suggested 
the need for a second parameter that could account for differences in HB morphology not obviously 
caused by metallicity. \citet{va65} summarized the problem, stating that metallicity is not 
sufficient to explain the extent of observed HB morphologies but that differences in age 
or He enrichment could explain the observed variations.  Both of these suggestions remain valid
up to the present time. \citet{va67} analyzed the integrated colors of 49 GCs 
with $UBV$ photometry and concluded that `at least two parameters (one of which is metal 
abundance) are required to describe globular clusters.' 

The second parameter phenomenon took on a greater significance with the 
seminal work of \citet{sz78}.  Searle \& Zinn recognized that GCs with 
unusually red HBs are relatively rare in the inner regions of the Galactic halo
(Galactocentric radius, $\rgc \la 8$ kpc) but become increasingly common at greater $\rgc$
(see their Figure 10). Searle \& Zinn used this fact to argue that the inner halo 
was assembled early and in a fairly short time while the outer halo was assembled over
an extended period of time.
Current galaxy formation scenarios envision the outer regions of the Galaxy as the accumulated 
debris of the many accretion events that shaped the early evolution of the Galaxy. In that context, 
understanding the origin and existence of age and metallicity gradients in the Galactic GC
population is as relevant now as it was at the time when Searle \& Zinn first introduced their
halo formation scenario.

Subsequent efforts relating to the second parameter 
problem fall into two general categories: those that attempt to measure the age difference 
between two (or a few) GCs with similar metallicities but markedly different HB morphologies 
and those that investigate the ensemble properties of a large sample of GCs.

One canonical pair in the former category is NGC~288, with a blue HB, and NGC~362,
with a red HB. The first CCD-based, differential photometric study of these clusters was 
conducted by \citet{bo89}. Bolte concluded that NGC~288 is $\sim$3 Gyr older than NGC~362. 
The subsequent work of \citet{gr90}, 
\citet{sa90}, and \citet{va90} echoed this result. By contrast, \citet{vd90} reached a very 
different conclusion. Using the brightest RGB star in each GC, 
\citet{vd90} corrected the photometry of NGC~288 and NGC~362 for their relative distances 
and found their main sequence turnoffs to be roughly coincident, thereby suggesting a negligible 
age difference. \citet{be01} found NGC~288 to be 2$\pm$1 Gyr older than NGC~362 using three
different techniques; in a follow-up, \citet{ca01} found that an age difference of 2 Gyr was
plausible if both GCs have $\feh \sim -1.2$ but that their synthetic HB models were unable to
match the detailed HB morphology of either GC using canonical assumptions of average mass loss
and dispersion on the HB. In considering differences between second parameter pairs, it is 
important to understand just how similar the abundances are in both clusters. \citet{sh00} 
performed a detailed comparison of abundances in red giants with 13 stars in NGC~288 and 12 in 
NGC~362. These authors concluded that NGC~288 has a lower $\feh$ than NGC~362 by 0.06 dex, 
that the average $\afe$ ratios are very similar, and that both GCs exhibit variations among 
O, Na, and Al.

The other canonical second parameter pair is M~13, with a blue HB, and M~3, with an intermediate 
HB.  M~13 and M~3 are $\sim$0.2 dex more metal poor than NGC~288 and NGC~362. 
\citet{va90} found evidence for an age difference but the quality 
of their CMDs did not allow a more definitive statement. \citet{ca95} estimated an upper limit 
to the age difference of $\sim$3 Gyr and concluded that it was insufficient to explain the HB 
morphologies, assuming both GCs have the same chemical composition. Alternatively, \citet{jo98} 
suggested the difference in their HB morphologies was due to a difference in their He abundances. 
Johnson \& Bolte suggested that M~13 had a He mass fraction $\sim$0.05 greater than M~3. However,
\citet{sw98} showed that a difference of $\Delta$Y $\sim 0.05$ at constant $\feh$ would make the 
level of the HB brighter by $\sim$0.2 mag. Such a difference between M~3 and M~13 was ruled out 
by the photometry of \citet{re01}, who also concluded that M~13 is older than M~3 by $1.7\pm0.7$ Gyr.
\citet{sn04} compared spectra of 28 red giants in M~3 with 35 in M~13 and found the two GCs' mean 
$\feh$ values the same within their 1-$\sigma$ error bars. However, \citet{sn04} reported differences 
in light element abundance variations, particularly O, for which M~13 displays a larger range of 
variation than M~3 by $\sim$0.5 dex.

A discussion of the search for the second parameter in GC-to-GC comparisons would not be 
complete without including the work of \citet{st99} and \citet{do08b}. These investigations 
used HST WFPC2 photometry to measure the ages of the outer halo GCs Palomar~3, Palomar~4, 
and Eridanus \citep{st99} and AM-1 and Palomar~14 \citep{do08b} relative to inner halo 
GCs with similar metallicities, M3 and M5\footnote{There is substantial evidence that M~5 is
actually an outer halo GC currently near its perigalacticon, see \citet{sc96} and \citet{di99}.}. 
Each of the five outer halo GCs has a redder HB morphology than its comparison inner halo GC. 
Both studies concluded that the outer halo GCs are $\sim$1.5-2 Gyr younger than M3 and M5, provided 
the chemical compositions of the outer halo GCs are comparable to their inner halo counterparts. 
A recent study by \citet{ko09} found the abundances of Pal~3 are essentially the same as inner halo 
GCs of similar metallicity and thus the relative age comparison with M~3 by \citet{st99} was justified. 
Overall, the outer halo GCs' chemical compositions remain poorly understood by comparison with the 
inner halo, see e.g.\,\citet{pvi05}. \citet{ca00} found the reported age differences between Pal~4 and 
Eridanus, with red HBs, and M~5, with an intermediate HB, too small to explain the difference
in HB morphologies unless all three GCs are younger than 10 Gyr, assuming standard assumptions of 
mass loss on the RGB.  \citet{ca01b} concluded that it was possible to explain the difference 
in HB morphology between Pal~3 and M~3 if the former has less HB mass dispersion and is younger 
than the latter. These examples indicate that if age is the second parameter, then it is 
our lack of understanding of mass loss that confuses GC-to-GC comparisons as first pointed out by 
\citet{ro73}. The review by \citet{ca05} includes a thorough discussion of different mass loss 
prescriptions and their efficacy. Given the HB morphologies and relatively young ages of the most 
distant outer halo GCs\footnote{The exception in the outer halo is NGC~2419. A number of dedicated 
HST photometric studies have focused on this massive, distant GC. \citet{ha97} and \citet{sa08} 
both found NGC~2419 to be an old, metal poor GC with a blue HB.}, it is important to 
include them when considering the properties of the entire Galactic GC population, especially in 
the context of Searle \& Zinn's halo formation scenario. 

The second category of second parameter studies 
are those that consider the properties of a large sample of Galactic GCs, e.g.\,\citet{sz78}.
In the first CCD-era study, \citet{sa89} compiled properties for 31 Galactic 
GCs. Among other things, \citet{sa89} examined the variation of HB morphology with age for 
GCs in a narrow range of metal abundance and showed that GCs with red HBs are significantly 
younger than those with blue HBs by as much as 5 Gyr. The subsequent 
studies of \citet{ch92}, \citet{sa95}, and \citet{ch96} updated and reaffirmed this 
result. \citet{ro99} found that the GCs with $\rgc > 8$ kpc are, on average, younger 
than those with $\rgc < 8$ kpc, which is consistent with age as the second parameter.
By contrast, \citet{ri96} examined the CMDs of 36 GCs, found an age dispersion of $\sim$1
Gyr with no significant age gradient in the Galactic halo, and concluded that this age
range was too small for metallicity and age alone to explain HB morphology in GCs.

Theoretical efforts have provided further insights into the complexities of HB morphology.
In particular, several studies have applied the synthetic HB model developed by \citet{ro73}.
\citet{ldz90}, \citet{ca93}, and \citet{ldz94} used synthetic HB models to explore the 
interplay of HB morphology, metallicity, and age in the CMD. These studies demonstrated 
considerable degeneracy in the HB morphology--metallicity diagram and, in particular, \citet{ca93}
argued that unless absolute values of the He abundance, $\afe$ ratios, and RGB mass loss were
known, HB morphology did not constitute a reliable age constraint.  Nevertheless, \citet{ldz94} 
concluded that there was evidence for an age dispersion of $\la 5$ Gyr among Galactic GCs of 
similar metallicity but markedly different HB morphologies. A number of 
theoretical studies have also focused on second parameter pairs or triads but, unfortunately, 
the lack of a firm theoretical understanding of mass loss along the RGB has impaired the these 
efforts, as first pointed out by \citet{ro73}. See \citet{ca05} for a recent review.

As it pertains to HB morphology, the use of the term 'second parameter' has expanded to cover 
a broad range of complex behaviors. In particular, this includes the faint extent of the blue 
HB tail in some GCs. \citet{fp93} analyzed 53 GCs and found that the length of the HB 
in the CMD--and the extent of the blue tail--is correlated with central density.  \citet{bu97} 
also reported a link between an extended blue tail and central density in GCs and concluded 
that `environment is ``a'' second parameter'.
\citet{sm04} showed that there is a correlation between central density and HB morphology 
for intermediate metallicity GCs ($-1.7 < \feh < -1.3$) where the second parameter effect
is most pronounced.
\citet{rb06} analyzed the properties of 54 Galactic GCs with homogeneous photometry from an 
HST WFPC2 Snapshot Survey \citep{pi02} and concluded that more massive GCs tend to have more 
extended blue HBs. \citet{rb06} recognized a link between the effective temperature of the 
hottest HB star in a GC and its mass (as inferred from its integrated luminosity) and suggested 
that self-pollution could explain the existence of faint blue tails in preferentially massive GCs. 

Evidence abounds for the presence of chemical abundance variations in all GCs \citep{gr04,car09a,car09b} 
and the possible correlation between the degree of abundance variations and the faint extent of the blue 
HB \citep{car07}. However, it is unlikely that the existence of faint, blue tails in the HBs of some GCs 
is directly related to the appearance of anomalously red HBs in others. It is also unlikely that chemical 
abundance variations will unduly affect age estimates of most GCs, provided the total metal content is 
constant across all stars \citep{pie09}. For cases in which the total metal abundance is not constant, 
or there are distinct stellar populations present in the CMD, age estimates are necessarily more complicated 
and careful analysis of each of these GCs is needed. See \citet{pi09} for a recent summary of GCs with 
multiple stellar populations visible in the CMD, several of which were discovered with photometry from the 
ACS Survey of Galactic GCs. The complex issue of multiple stellar populations in GCs remains poorly understood 
and the extent to which multiple-population GCs permeate the Galactic GC population is unknown at present.

To summarize, the presence of metal poor GCs with red HBs that primarily reside in the outer 
Galactic halo is well-known observationally. Although the present study is focused on the Galactic GCs,
including some of those associated with the Sagittarius dwarf, there is ample evidence to suggest that
metal poor GCs with red HBs are also found in the Magellanic Clouds \citep{jo99,gl08} and Fornax \citep{bu99}.
\citet{mg04} summarize our current knowledge of the GC populations in the Galaxy and its satellites
in the HB morphology--metallicity diagram.
Despite an abundance of observational evidence, no consensus has been reached as to 
what parameter(s) are responsible for the appearance of relatively red HBs in metal poor GCs. 
Age is frequently offered as a candidate but considerable doubt still remains because the age 
difference claimed--or required by theoretical studies of HB morphology--is too large to satisfy the 
observations.

The existence of a homogeneous database of deep, high quality photometry from the ACS Survey 
of Galactic GCs \citep[Paper I in this series]{sa07} has motivated a re-examination of HB morphology 
and its relation to a variety of GC properties. The paper is organized as follows: 
$\S$\ref{data} describes the data sample;  $\S$\ref{hbm} describes the methods that were employed to 
determine the HB morphologies; $\S$\ref{gcages} describes the sources of GC ages and provides some
discussion of complicating factors in age determination; $\S$\ref{results} presents the analysis 
performed on the assembled data and discusses the important results; and, finally, $\S$\ref{conclusion}
provides a summary of the salient points.

\section{The observational data\label{data}}

\subsection{The ACS Survey of Galactic GCs\label{survey}}
The photometric catalog of the ACS Survey of Galactic GCs \citep[Paper V]{an08} consists of
65 GCs observed in $F606W$ and $F814W$. Most details concerning the data reduction and calibration 
are provided in Paper V but, since its publication, the data have been adjusted to account for 
updated HST/ACS Wide Field Channel zero-points \citep{bo07}; the new $F606W$ and $F814W$ zeropoints 
are, respectively, 22 and 35 mmag fainter than given by \citet{si05}.

Not all GCs in the ACS Survey catalog will be considered in the analysis performed in the
following sections.  $\omega$ Cen (NGC~5139) and M~54 (NGC~6715) have been excluded 
because their complex CMDs and abundances indicate that these objects are not GCs in the traditional 
sense. \citet{si07} analyzed the M~54 CMD and measured ages for the stellar populations present 
and also estimated the metallicities of the stellar populations for which no spectroscopic 
information was available. Because M~54 lies at the center of the Sagittarius dwarf galaxy, it is 
not a simple process to distinguish between the stellar population(s) that belong to M~54 and those 
of the Sagittarius dwarf galaxy.  $\omega$ Cen has long been known to contain a range of different 
metallicities as well as multiple stellar populations in the CMD \citep{le99}. \citet{jo08} presented 
a spectroscopic analysis of 180 red giants in $\omega$ Cen that revealed at least four distinct 
metallicities and a total range of $\sim1.5$ dex in $\feh$. Other GCs for which multiple stellar 
population evidence exists will be noted in the following sections. \citet{pi09} provided a summary 
of the GCs known to exhibit multiple stellar populations in their CMDs circa 2008.
Three more GCs were excluded: Pal~1 and E~3 for lack of identifiable HB stars and Pal~2 due to extreme 
differential reddening (see Paper I). In total, 60 GCs from the ACS Survey are considered in the following analysis.

\subsection{The six most distant GCs}
The ACS Survey provides coverage of the Galactic GC population out to $\sim$20 kpc.  
In order to give proper consideration to the outer Galactic halo, where the second parameter 
phenomenon is most pronounced, the sample was extended to include the following GCs: 
NGC~2419 \citep{ha97,sa08}; Pal~3, Pal~4, and Eridanus 
\citep{st99}; and AM-1 and Pal~14 \citep{do08b}. These are the six most distant GCs 
in the Galaxy with $70 \la \rgc \la 120$ kpc \citep{ha96}.  Unfortunately, no photometry 
of comparable quality is currently available for the 10 or so GCs that lie between these 
two groups ($ 25 \la \rgc \la 50$ kpc)\footnote{Pal~2 is one of these, and part of the ACS
Survey catalog, but its CMD is so obscured by differential reddening that no useful information 
can be extracted from it (see Paper I). We note that during HST Cycle 17, GO 11586 will use ACS 
to image three GCs at $\rgc \sim 40$ kpc: Pal~15, NGC~7006, and Pyxis.}. In order to
distinguish between the homogeneous ACS Survey sample and the larger, heterogeneous sample, 
the statistical analysis in section $\S$\ref{results} will be presented with and without the 
six outer halo GCs.

\subsection{Additional GC parameters}
In addition to HB morphologies ($\S$\ref{hbm}) and ages ($\S$\ref{gcages}), several GC parameters 
were extracted from \citet{ma05}. The \citet{ma05} catalog is based on the \citet{ha96} catalog 
but includes several updated quantities and, importantly, provides quantities in absolute physical
units. These units are more appropriate for comparisons than the observational units given in the
\citet{ha96} catalog.  The quantities obtained from the \citet{ma05} catalog are:
Galactocentric distance ($\rgc$), integrated absolute $V$ magnitude ($\mv$), half-light radius ($\rh$), 
and tidal radius ($\rt$). From the \citet{ha96} catalog we extracted central luminosity density ($\rho_0$). 
Neither catalog includes basic measurements for Lyng\aa~7. Inclusion of these quantities in the following 
analysis makes it possible to explore the possibility that a GC's location in the Galaxy, structural 
properties, and total mass (assuming the mass-to-light ratio is constant) can influence its HB morphology.

\section{Measuring HB morphology\label{hbm}}

In this study, HB morphology is characterized by the difference between 
the median color of the HB stars and the median color of the RGB at the level of the HB, 
denoted $\dvi$. This metric is essentially the same as $d_{B-V}$ \citep{sa99}
and is less likely to become saturated than the most frequently used HB morphology parameter, 
(B$-$R)/(B+V+R) introduced by \citet{lee89}.
As a corollary, the magnitude level of the HB in $F606W$ has also been measured; it provides a
reference for GC distance estimates as discussed below.

\subsection{The mean magnitude level of the HB\label{vhb}}
The mean level of the horizontal part of the HB was measured from the ACS Survey data 
relative to the HB of M~5 (NGC~5904). M~5 has a well-populated HB that extends approximately 0.7 mag in $F606W-F814W$
from the red side to the blue and, therefore, can overlay the HB of almost any other GC.
Measurements were made by matching the CMD of each GC to that of M~5 in the region of the HB by
making vertical and horizontal adjustments to the comparison CMD until it most
closely overlaid that of M~5. In the majority of cases, the comparison GC had enough overlap
to make the best alignment unambiguous.  A few clusters with purely blue HBs, such as
NGC~6254, present some difficulty because even the reddest HB stars do not become `horizontal'.
The measured HB levels are listed in Table \ref{morph}. Typical measurement uncertainties are
$\sim$0.05 mag and are dominated by the uncertainty inherent in aligning the HBs. The formal 
photometric errors of most HB stars are a few thousandths of a magnitude.
The uncertainty is larger in cases where differential reddening is significant 
or the HB is otherwise difficult to quantify.
As an example, Figure \ref{outline} shows HB levels for NGC~1261 and M~92 (NGC~6341).

\subsection{The color difference, $\dvi$\label{deltavi}}

\begin{figure}
\plotone{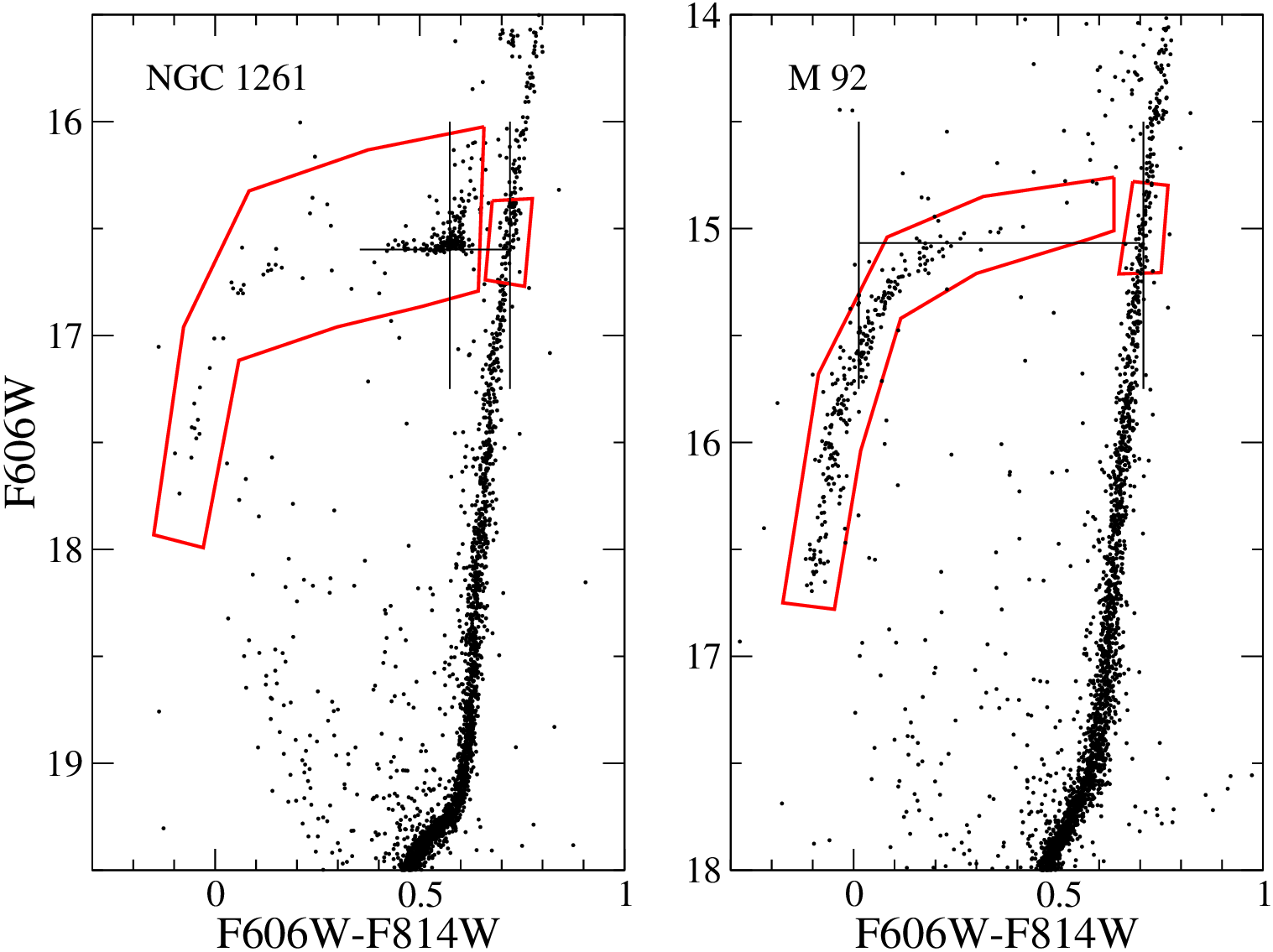}
\caption{Portions of NGC~1261 and M~92 (NGC~6341) CMDs are shown. Thick lines indicate the regions that were
selected to determine the median HB colors.  Thin lines indicate measured quantities: vertical
lines indicate the median colors of the HB and RGB; the horizontal line indicates the level of
the HB.\label{outline}}
\end{figure}

Measurements were obtained by drawing an outline around the HB and the RGB at the 
level of the HB and then calculating the statistics of the stars within each 
outline. This process is illustrated in Figure \ref{outline} for two GCs, NGC~1261 and M~92.
RR Lyrae stars are not given special consideration in this analysis: they are included if
they are present in the ACS observations. For the ACS Survey data, all determinations of HB membership 
were performed in the ACS $F606W-F814W$ CMD as shown in Figure \ref{outline}.
In order to allow direct comparison with GCs observed in different bands, the color difference
presented in Table \ref{morph} is given in terms of $V-I$ rather than $F606W-F814W$. 
This was achieved by transforming the ACS magnitudes of each individual HB star into $V$ and $I$ 
using the equations provided by \citet{si05} and then calculating the median $V-I$ values of the
HB and RGB. The median was chosen to represent 
the central tendency of $\dvi$ because the HB stars are not normally distributed in color space. 
Hence, the more commonly-used mean and standard deviation are not the most appropriate choices. 
If the HB stars were normally distributed in color space, then the mean and median would be equal. 
In fact, about one half (one quarter) of the GCs in the sample have means and medians 
that differ by more than 5\% (10\%). For GCs with intermediate HBs, such as M~3 and NGC~1261, 
the mean and median are substantially different.

Table \ref{morph} provides the quantities used to determine $\dvi$. 
In addition to the median $V-I$ of the HB and RGB, 1-$\sigma$
errors on these quantities are listed and the quadrature sum is given as the uncertainty
on $\dvi$. Uncertainty in the median $V-I$ was estimated by bootstrapping with replacements 
performed 10,000 times on both the HB and RGB of each GC.  The error bars on the median represent 
the range of $V-I$ within which 68\% of bootstrapped medians lie. To compliment the median and 
uncertainty, we also give the mean absolute deviation (MAD) for the HB and RGB.  The mean absolute 
deviation represents the intrinsic spread in each individual data set. Consider, for example, the 
HBs of 47~Tuc (NGC~104) and M~3 (NGC~5272). Both have well-populated HBs and thus small 
uncertainties, $\sigma < 0.01$ determined from bootstrapping, but 47~Tuc has a tightly clustered 
red clump of HB stars and a small MAD=0.023 while M~3 has a more broadly distributed HB and a
larger MAD=0.208.

\subsection{Comparison with other HB morphology parameters}

\begin{figure}
\plotone{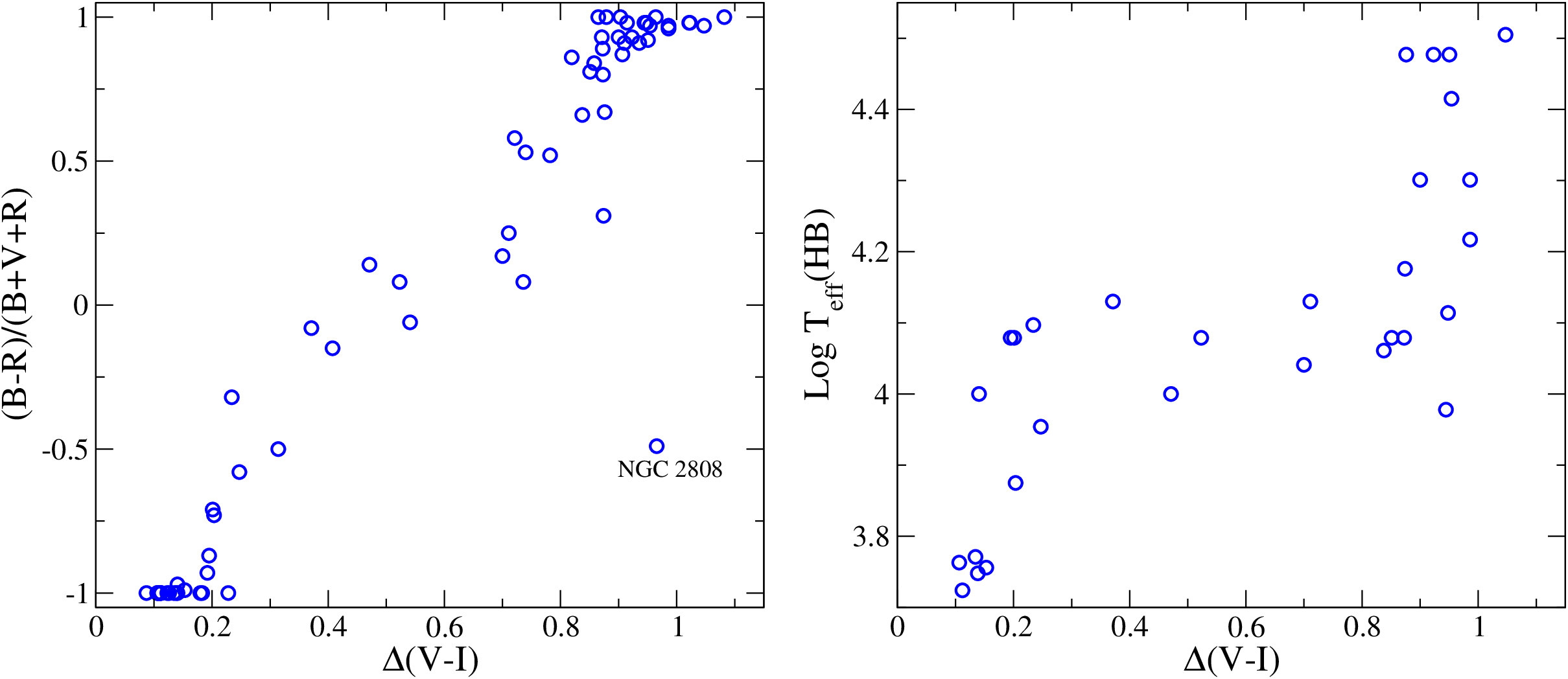}
\caption{{\it Left:} Comparison of $\dvi$ from this paper with (B$-$R)/(B+V+R) from 
\citet{ma05} for all 66 GCs in the sample. {\it Right:} Comparison of 
$\dvi$ and Log~$\thb$ from \citet{rb06} for the 30 GCs common to both studies.
\label{dVIBVR}}
\end{figure}

Perhaps the most frequently used HB morphology metric in the literature is (B$-$R)/(B+V+R):
the difference between the number of blue HB stars (B) and the number of red (R) normalized 
by the total number of HB stars including variables (V). Its major inconvenience is that it
saturates if all stars lie on one side or the other of the instability strip. The left panel 
of Figure \ref{dVIBVR} compares (B$-$R)/(B+V+R) from \citet{ma05} and $\dvi$ from 
this paper for the 66 GCs in the present sample. There is an obvious correlation between the 
two but $\dvi$ continues to vary after (B$-$R)/(B+V+R) saturates.

The most recent large-scale study of HB morphology, that of \citet{rb06}, used the maximum 
effective temperature encountered along the HB [Log~$\thb$]
determined with the use of theoretical zero-age HB sequences from \citet{ca99}. The Log~$\thb$ 
metric appears to be complimentary to $\dvi$ or (B$-$R)/(B+V+R) because, while it lacks 
sensitivity through the middle, it is more sensitive at the extremes. This is demonstrated 
in the right panel of Figure \ref{dVIBVR}. In the absence of an ideal HB morphology metric 
that quantifies every feature of interest, it is beneficial to identify a metric that is 
well suited to a particular problem.

\section{Globular cluster ages\label{gcages}}

\subsection{Relative ages\label{relage}}
\citet[Paper VII]{mf09} presented relative ages measured from the ACS Survey data. These ages
were derived by first measuring the absolute magnitudes of the main sequence turnoffs (MSTOs)
for all GCs and then interpolating in  
isochrone-based grids of MSTO as a function of age and metallicity. The same analysis was
performed with four different isochrone libraries and the distribution of relative ages with
metallicity was shown to be independent of the isochrone library. For the purpose of the present paper, 
relative ages derived from \citet[Paper II]{do07} isochrones using the \citet[hereafter ZW84]{zw84}
metallicity scale have been adopted. Furthermore, we assumed that GCs with $ \feh < -1$ have
[$\alpha$/Fe]=+0.3 and GCs with $ \feh \geq -1$ have [$\alpha$/Fe]=+0.2.  The ages measured
in this fashion and with these assumptions were placed on a relative scale by dividing out the 
average age of the most metal-poor GCs ([Fe/H] $< -1.8$), in this case 13.3 Gyr. In the present study, 
this factor has been retained, resulting in absolute ages for the particular isochrone library, 
metallicity scale, and method of age determination employed in Paper VII.

\subsection{Isochrone fitting\label{isoage}}
In order to put the ACS Survey ages on the same scale as the six outer halo GCs, and to demonstrate
that later results do not depend on the method used, age estimates were determined using 
isochrone fitting to the CMDs of all GCs in the data set.  The fits were performed using isochrones from
Paper II for the ACS Survey data and \citet{do08a} for the outer halo GCs. The same zeropoint 
corrections applied to the ACS data, as described in $\S$\ref{survey}, were applied to the isochrones 
in the ACS photometric system as well. The underlying 
luminosities, temperatures, surface gravities, and color transformations of the isochrones in 
Paper II and \citet{do08a} are identical.  The only difference between the two is the photometric system.  
The outer halo GCs were measured in $F555W$ and $F814W$, whereas the ACS Survey data were taken in 
$F606W$ and $F814W$.  Several GCs were excluded from isochrone fitting because they are known to
harbor multiple stellar populations.  The presence of more than one stellar population
makes a single age determination insufficient to characterize the GC.  Ages were not determined
for the following GCs: NGC~1851 \citep[Paper III]{mi08}, NGC~2808 \citep{pi07}, NGC~6388, 
NGC~6441\footnote{Given the great similarities between the CMDs of NGC~6388 and NGC~6441 we have chosen
to exclude both although \citet{pi09} only showed evidence for multiple populations in NGC~6388.}, 
and NGC~6656 \citep{pi09}.

It is likely that other GCs with multiple populations exist, even within the ACS Survey data, but to have
avoided detection thus far the separation between the populations must either be so small that they overlap 
to a great extent or that reddening has obscured the separation in the CMD. 
A second population may also represent a small fraction of stars and have little 
or no apparent influence on the HB morphology. The recent discovery of a second subgiant branch 
in 47~Tuc (NGC~104) by \citet{an09} is an example: this second population accounts for
only $\sim$10\% of the stars in the core. Further out, the CMD was too sparsely populated to detect
this population.  It should also be noted that \citet{an09} found 
the main population of 47~Tuc to have a spread in the CMD greater than can be accounted for by the
photometric errors.  The authors found it unlikely that this spread was due to binaries, differential
reddening, or a depth effect. \citet{an09} explored the possibility that the color spread could
be caused by an intrinsic variation in the He abundance or $\feh$. Another case is M~4 (NGC~6121). 
\citet{ma08} presented spectroscopic evidence for a bimodal abundance distribution along the RGB of 
M~4 as well as a broadening of the RGB in $U-B$. Whether M~4 contains two distinct stellar populations 
or a continuous distribution is not clear at present.  If it is the former, the difference is small 
enough to remain undetected on the main sequence in the exquisite photometry presented by \citet{be09}. 
If it is the latter, age determinations are expected to be reasonable (see the discussion at the end of
 $\S$\ref{metage}).

The procedure used to determine ages involved a two step process. Initial estimates for $\feh$, 
distance modulus, and reddening estimates were taken from the \citet{ha96} catalog, 2003 revision. 
Estimates of [$\alpha$/Fe] were chosen as appropriate for a given value of $\feh$.  All efforts 
were made to follow these values as closely as possible with minor adjustments made to improve the 
fit to the unevolved main sequence first and the RGB second.  In some instances, isochrones 
at the catalog values were not able to provide an adequate fit to the CMDs and, in these cases,
the initial estimates of $\feh$, distance, and reddening were allowed to vary until the fit
obtained was acceptable.  This step was a necessity in a number of cases because the published
values for some lesser-known GCs are uncertain. 

For example, \citet{do08b} found that the main sequence and RGB morphologies of AM-1 are comparable 
to those of M~3 although the former is listed at 
$\feh=-1.8$ while the latter is listed at $\feh=-1.57$ in the Harris catalog. Another example
is the distance modulus of NGC~6254: the Harris catalog value is $DM_V=14.08$ but the present study
produces a value that is $\sim$0.2 mag fainter. This result is confirmed by both isochrone
fitting to the main sequence and the level of the HB from Table \ref{morph}. These results are not 
meant to diminish the value of the GC catalogs, which are invaluable resources, but merely to emphasize
that there are still many GCs in the Galaxy whose properties are poorly constrained.  The models,
which are homogeneous in terms of physical assumptions and ingredients, can prove useful
in determining the relative differences between two GCs if one has well-measured properties
and another does not.

Several of the GCs in the ACS Survey are heavily obscured by interstellar reddening.  
Differential reddening confuses the age determinations in such GCs because the reddening 
line is nearly perpendicular to the most age-sensitive features: the main sequence turnoff 
and the subgiant branch. Differential reddening causes these features to be spread out and, 
in light of this trend, we have attempted to correct for its effects in the following manner. 
The first step involves the construction of a fiducial sequence for the GC. The fiducial 
sequence is centered approximately at the MSTO. More stars lie below the MSTO than above but 
the differential corrections are weighted based on the angle between the stellar sequence
and the reddening line. Since the reddening line is almost perpendicular to the subgiant
branch, the differential correction is approximately equally weighted by stars above and
below the MSTO. Care is taken to avoid unresolved binaries.
From the fiducial sequence, each individual star yields a color residual, taken along the 
reddening direction in the CMD. From these residuals a reddening map is created by finding 
the median residual in each $256\times256$ pixel square of the image. Then each star is 
corrected, along its reddening line, by an amount that is interpolated from the $16\times16$ 
points of the reddening map. A full explanation of these procedures, the 
differentially corrected CMDs, and reddening maps will be provided in a forthcoming 
paper (I. King, 2009, in preparation).

The best-fit age of a given GC was estimated by determining the isochrone
that best fit the CMD from the MSTO through the subgiant branch. The uncertainty was derived
from the intrinsic scatter in the CMD and/or the inherent mismatch between the models and the data.
The range of ages that allowed the isochrones to envelope the bulk of the stars 
at the MSTO and on the subgiant branch are taken to be 1-$\sigma$ uncertainties.
The age uncertainty is only based on the fitting procedure described and does not account 
for uncertainties in the input physics of the stellar models or differences between the 
chemical composition assumed in the models and those actually present in the stars  
(for more on this point, see $\S$\ref{metage}). Hence the uncertainties include the random component 
but exclude the systematic. Still, it is worthwhile to consider that incomplete knowledge of chemical
composition and incomplete treatment of the physics, such as rotation and convection, in the current
generation of 1-D stellar evolution models are primary sources for the systematic error present in the
analysis.

\begin{figure}
\plotone{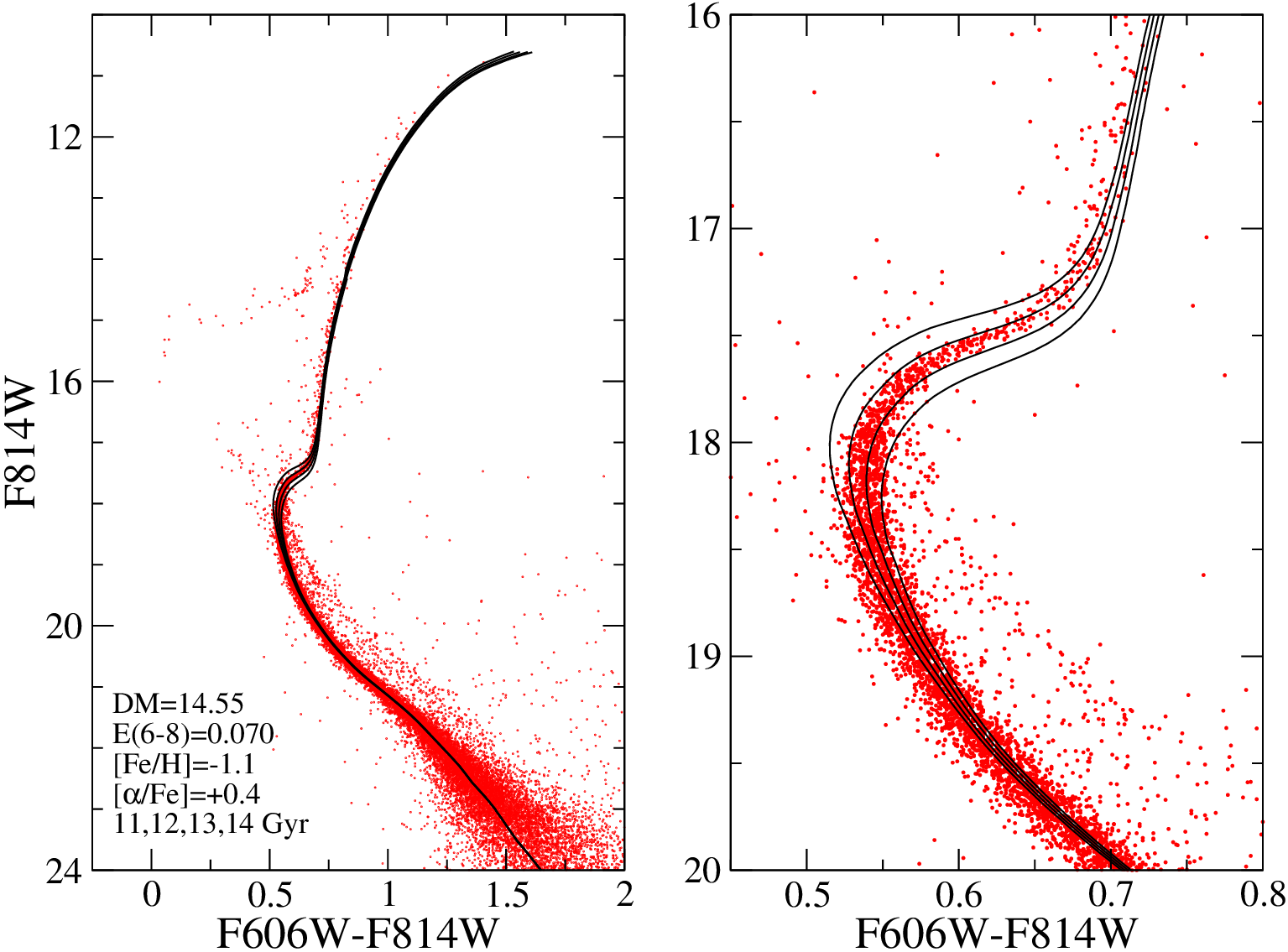}
\caption{Isochrone fits to the CMD of NGC~6362; the fit parameters are listed on the figure.
The left panel shows the full CMD to indicate the agreement between the models and the data
on the RGB and unevolved main sequence.  The right panel focuses on the main sequence turnoff
region to indicate how the age was determined.\label{iso6362}}
\end{figure}

\begin{figure}
\plotone{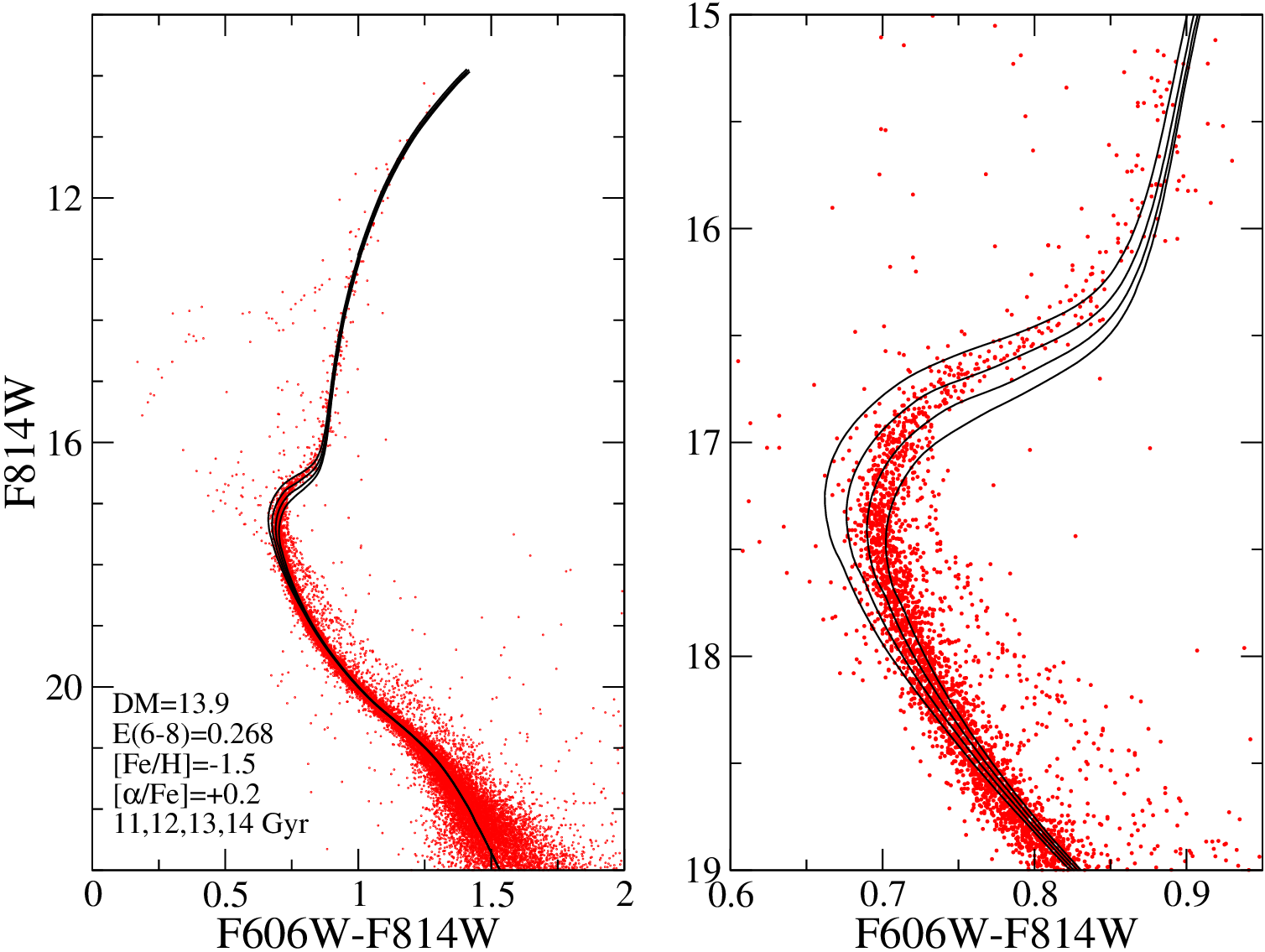}
\caption{Same as Figure \ref{iso6362} but showing the differential reddening corrected CMD of NGC~3201.\label{iso3201}}
\end{figure}

\begin{figure}
\plotone{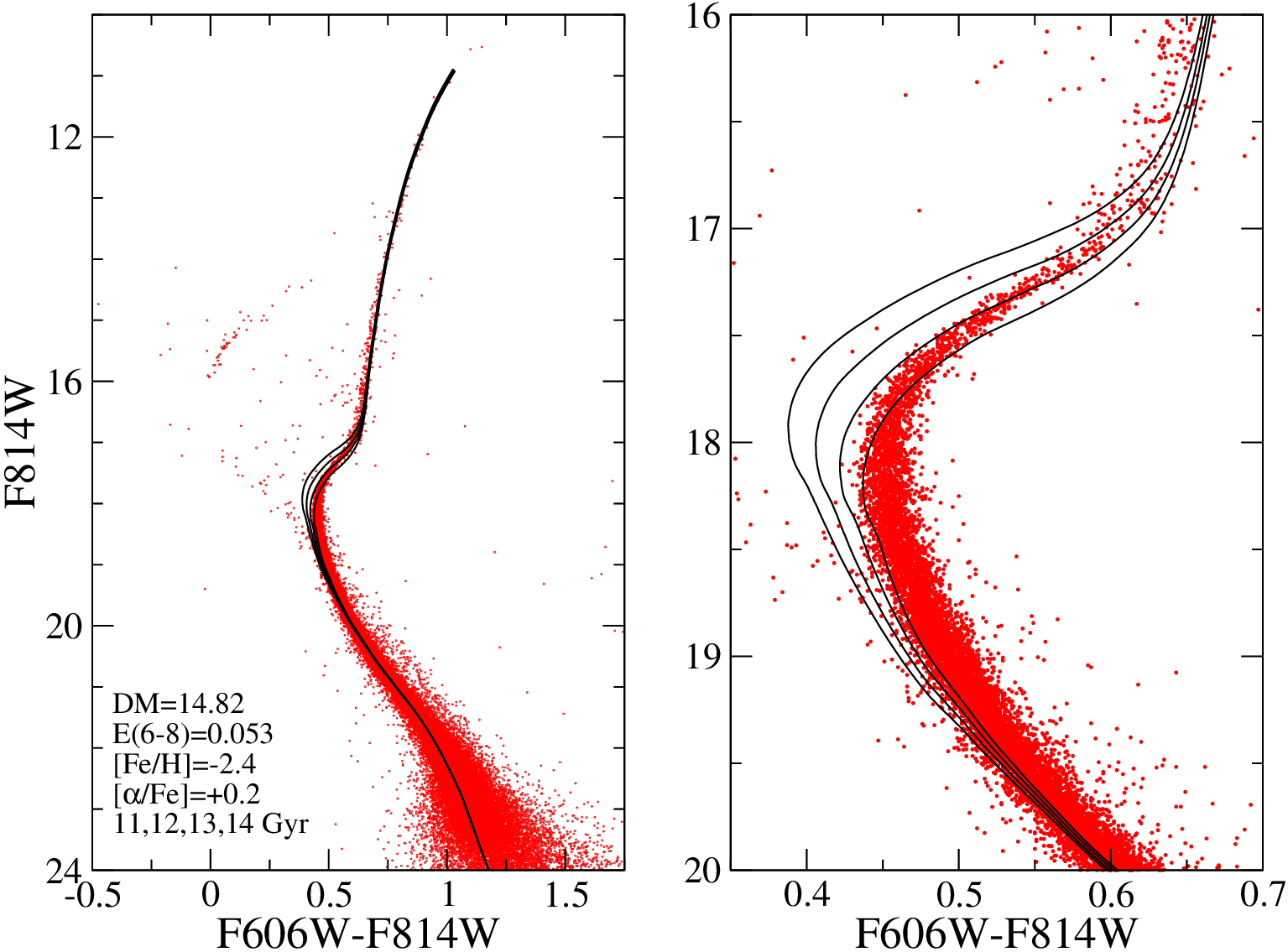}
\caption{Same as Figure \ref{iso6362} but showing NGC~7099.\label{iso7099}}
\end{figure}

The results of isochrone fitting are listed in Table \ref{age}.  If a differential reddening
corrected CMD was employed in the fit, an asterisk (*) appears after the name in Table \ref{age}.
Figures \ref{iso6362} through \ref{iso7099} demonstrate how the isochrone fits were achieved for 
three different GCs.  Figure \ref{iso6362}
shows a case where the isochrones trace the stellar population throughout the CMD
(left panel) and how the age, 12.5$\pm$0.5 Gyr, was measured (right panel).  Figure \ref{iso3201}
shows the differential reddening-corrected CMD of NGC~3201.  The
corrected CMD reduces the scatter about the subgiant branch but the age uncertainty is still larger
than in the case of NGC~6362. Figure \ref{iso7099} shows the worst case scenario where the 
reddening is low and the isochrones match the unevolved main sequence and RGB but not the shape of 
the age sensitive region.  The slope of the models is shallower than the data along the subgiant 
branch and, although the data define a narrow sequence, the age is poorly constrained. This case is
likely to be severely affected by the systematic errors described in the preceding paragraph.

\begin{figure}
\plotone{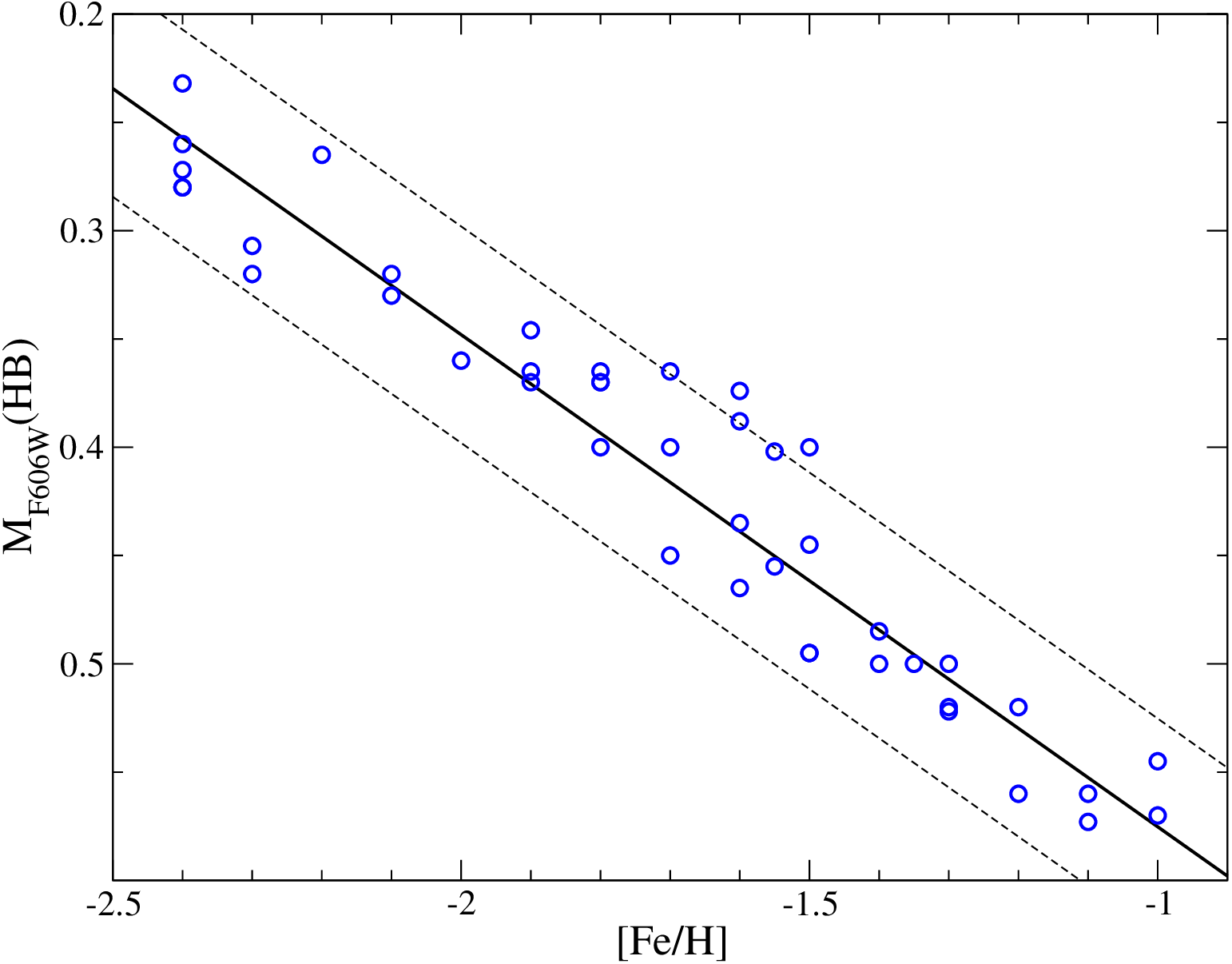}
\caption{Absolute magnitude of the HB in $F606W$ as a function of metallicity.  The best fit line
is shown as a solid line.  The dashed lines indicate $\pm0.05$ mag above and below the best fit line, 
which corresponds to the estimated uncertainty in the apparent magnitude of the HB as mentioned in 
$\S$\ref{vhb}.\label{mvhb}}
\end{figure}

As a consistency check, it is useful to plot the absolute magnitude of the HB as a function of metallicity. 
 The absolute magnitude of the HB is obtained by subtracting the $F606W$ distance modulus derived from the 
isochrone fits (Table \ref{age}) from the apparent magnitude of the HB (Table \ref{morph}).  
The resulting quantity is plotted in Figure \ref{mvhb} for the ACS Survey
clusters. The points define a relatively tight relationship. A linear, 
least squares fit to these data for GCs with $\feh \leq -1$ gives: 
\begin{equation}\label{mveq}
<M_{F606W}(HB)> = (0.227\pm0.011)\feh +  0.802\pm0.020,
\end{equation}
which is shown as the solid line in Figure \ref{mvhb}.
The quoted errors in equation (\ref{mveq}) are due only to scatter in the data and do not include measurement 
uncertainties.  The fit is limited to clusters with $\feh \leq -1$ so as to exclude clusters with predominantly 
red HB morphologies. The justification for restricting the fit is that red HB GCs often lack RR Lyrae variable 
stars. Some GCs with completely blue HBs were included in the fit even though they may not have RR Lyrae stars. 
As described in $\S$\ref{vhb}, the uncertainty in the apparent magnitude of the HB is approximately 0.05 mag per 
GC and this is shown by the two dashed lines. All but a handful of points lie within the dashed lines.

If equation (\ref{mveq}) is transformed from $F606W$ to $V$ using the synthetic color transformations
employed in Paper II, the slope increases by $\la0.01$ because of a very slight metallicity dependence
of $V-F606W$ on $\feh$ and, assuming a characteristic temperature for RR Lyrae stars of LogT=3.83, the
intercept increases by 0.09 mag. \citet{ch99} gave $M_V(RR)$ = (0.23$\pm$0.04)($\feh$+1.6) + (0.56$\pm$0.12).
Other recent estimates include $M_V(HB)$=(0.22$\pm$0.05)($\feh$+1.5)+(0.56$\pm$0.07) \citep{gr03} and
a further refinement of the slope to 0.214$\pm$0.047 mag/dex \citep{gr04a}.
If equation (\ref{mveq}) is transformed again so that the independent variable is ($\feh+1.6$) and the dependent
variable is $M_V$ rather than $M_{F606W}$, it becomes $M_V(HB) = 0.235 (\feh+1.6) + 0.53$ where the
error bars have been omitted for brevity.  The relationship between $M_V(HB)$ and $\feh$ derived here is
within the 1-$\sigma$ uncertainties from each of the three determinations previously mentioned.

\subsection{Metallicity scales and chemical abundance variations\label{metage}}

\begin{figure}
\plotone{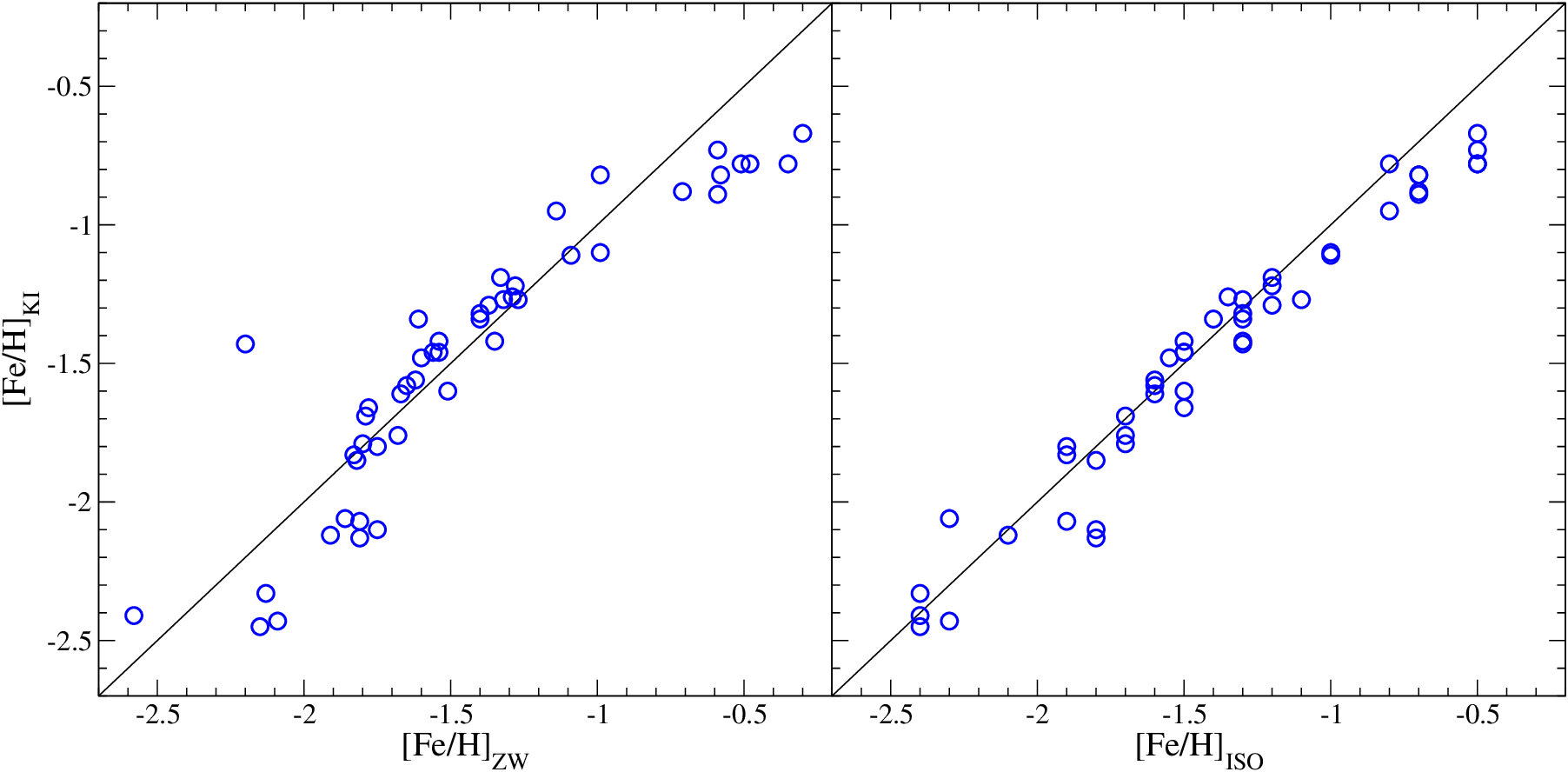}
\caption{Comparison of the $\feh$ scales from Table \ref{age} and ZW84 with the KI03 scale for 47 GCs.
The horizontal and vertical scales are the same in both panels.  The solid line in each panel indicates
equality.\label{met}}
\end{figure}

Figure \ref{met} compares $\feh$ from Table \ref{age} and the ZW84 scale with the  
\citet[hereafter KI03]{ki03} scale for 47 GCs in common between KI03 and this paper. (KI03 provided $\feh$
values measured with three different model atmospheres; this paper adopts their measurements based
on MARCS models.)
For reference, the line of equality is drawn on both panels of Figure \ref{met}. Paper VII 
compared ages on the ZW84 and \citet[hereafter CG97]{cg97} scales. As discussed by KI03, the CG97 
scale is consistently higher than either the ZW84 or KI03 scales, by 0.2-0.3 dex in general. It is beyond 
the scope of this paper to determine which scale is superior to the others\footnote{
Recently, \citet{car09c} presented a new GC metallicity scale based on
the largest spectroscopic survey of GC red giants to date. The analysis reveals a very close 
agreement between the new metallicity scale and that of KI03. This is in spite of 
the fact that KI03 used Fe\,II lines while \citet{car09c} relied on Fe\,I lines. 
On the other hand, the new scale is consistently $\sim$0.2 dex lower than that of CG97.}.
Suffice it to say that ages and metallicities from Table \ref{age} and Paper VII on the ZW84 scale will be 
used in the analysis that follows. However, speaking hypothetically, if it is assumed that the KI03 scale is 
the `correct' $\feh$ scale, then the residuals from the solid line in each panel of Figure \ref{met} can be 
interpreted as uncertainties in the adopted scales.  We shall return to this hypothesis in $\S$\ref{2ndP}.

The ages reported in Tables \ref{gcages} are based on isochrones that assume an $\alpha$-enhanced composition.
They do not explicitly account for variations in the He or C+N+O abundances but, if such variations are present,
they should certainly influence age determinations \citep{ven09,dan09}. While the extent of He and C+N+O variations
in GCs are largely unknown at present, the existence of light element abundance variations is quite clear.
The review by \citet{gr04}, the recent results presented by \citet{car09a,car09b}, and many other studies
have shown the widespread presence of chemical abundance variations within individual GCs.  These abundance 
variations are among the light elements (C, N, O, Na, Mg, and Al) and do not extend to heavier elements 
such as Fe in the majority of GCs. Models for GC self-enrichment predict that varying degrees of He variation 
should accompany the other abundance variations \citep{dan02,dec07}. The influence of such abundance variations 
on GC ages is not fully understood at present.  However, the recent results from \citet{car09a,car09b} lend some 
insight into the problem of deriving GC ages in the presence of light element abundance variations. Those authors
identify three groups of stars within GCs based on their chemistry.  They are: the `primordial' stars
or stars which have halo-like abundances, characterized by super-solar O and sub-solar Na; the 
`intermediate' stars, with slightly less O and more Na than the primordial stars; and the `extreme' stars, 
characterized by [O/Na] $< -$0.9. \citet{car09a,car09b} state that all GCs (measured so far) contain primordial 
and intermediate stars but not all GCs contain the extreme component.  Furthermore, the majority of GC stars 
are in the intermediate group.

How do these findings influence GC age estimates?   \citet{pie09} compared stellar evolution models
from the BaSTI library \citep{pie04,pie06} with chemical compositions representative of the primordial 
and extreme groups.  \citet{pie09} state that, as long as the total amount
of C+N+O remains constant, it is safe to use $\alpha$-enhanced isochrones to derive GC ages.  However,
if the total amount of C+N+O varies among the stars in a given GC, then it is necessary to use models
with the appropriate CNO abundances to derive GC ages.  This is all without consideration of a potential
change in the He content that is predicted to accompany the light element abundance variations. 
The He abundance is important because models of chemical enrichment in GCs
predict that the abundance variations from the primordial values should be accompanied by an increase in 
He, whether the source of the enrichment is massive stars \citep{dec07} or intermediate mass AGB stars 
\citep{dan02}. The extreme stars should be most He-enriched, the 
intermediate stars should be slightly He-enriched, and the primordial stars should have primordial He.  
\citet{sal04} investigated the He content of 57 GCs using the R parameter: the ratio of
the number of HB stars to the number of RGB stars brighter than the HB.  Within the uncertainties, they 
found little or no evidence for a spread in the He content of GCs with (B$-$R)/(B+V+R) $<$ 0.8.
For GCs with (B$-$R)/(B+V+R) $\geq$ 0.8, \citet{sal04} found a larger spread and a tendency toward 
higher He abundances. These authors state that the apparent trend towards higher He abundances in GCs with 
blue HBs may be related to increased evolutionary timescales for the lowest mass HB stars
not entirely accounted for in their calibration of the R parameter or the genuine presence of He-rich 
stars in the blue HBs of some GCs. 
As long as the total metal content remains constant, a modest spread ($\Delta$Y $\la$ 0.05) 
in the He content of a GC will not significantly alter the level of the main 
sequence turnoff or subgiant branch \citep{do09} and 
therefore not confuse the age determination. If the total metal content
varies or the spread in He is larger, age estimates will require great care and 
detailed abundance information.
It is important to keep in mind the complexities of measuring GC ages in the context of chemical abundance 
variations--and their proposed origins--in GCs.

\subsection{Age-Metallicity Relations}

\begin{figure}
\plotone{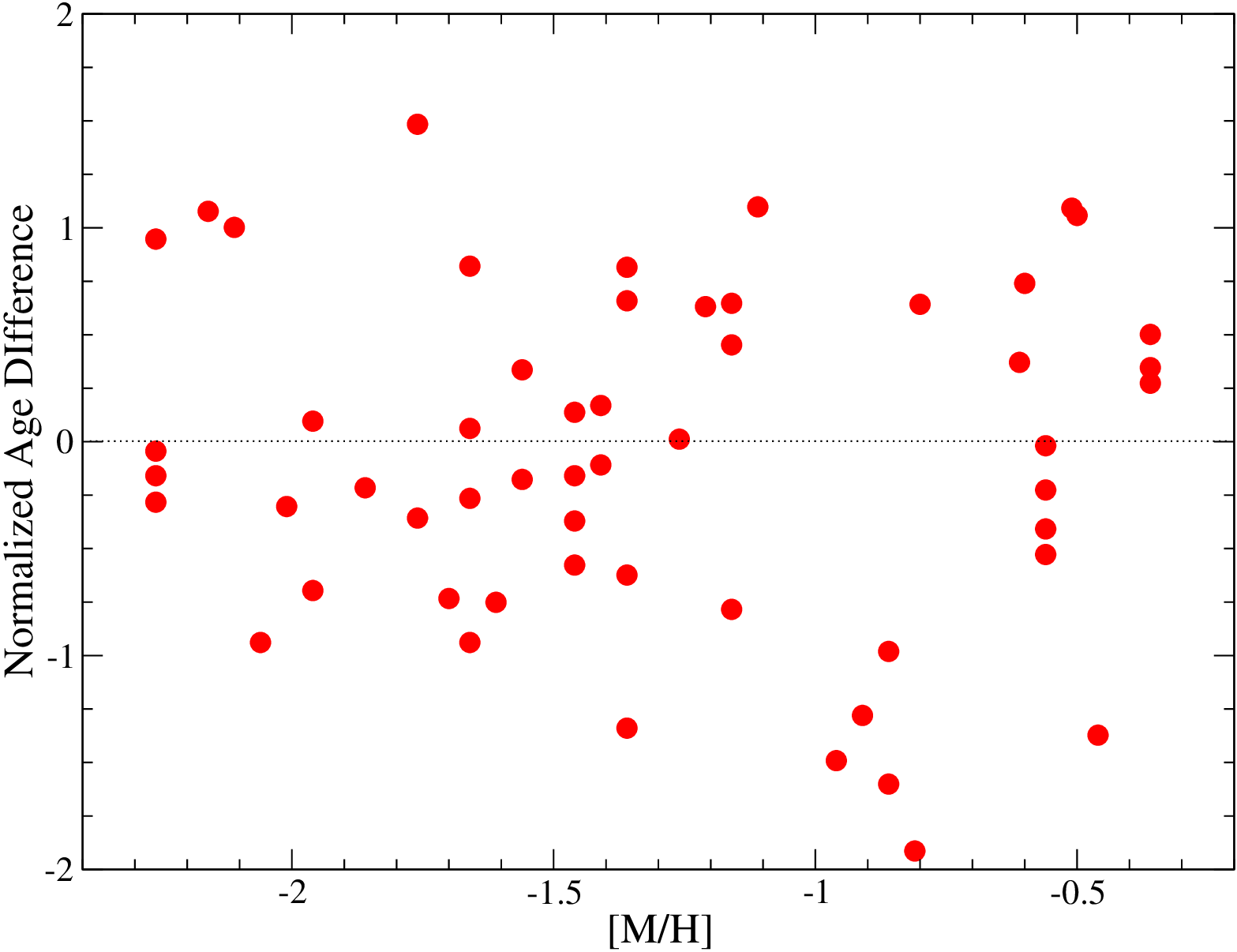}
\caption{The normalized difference in age between this study and Paper VII as a function of metallicity.
Age difference is presented in the sense that $\Delta$Age = Age(ThisPaper)$-$Age(PaperVII)
assuming, for consistency, ages from the latter based on the isochrones from Paper II and the ZW84 
metallicity scale and normalized by dividing by the quadrature sum of the age uncertainties from
both sources.\label{RelAges}}
\end{figure}

The average difference between the ages derived in $\S$\ref{isoage} and in Paper VII is $-0.104$ Gyr;
the standard error of the mean is 0.105 Gyr and the standard deviation is 0.781 Gyr. Figure \ref{RelAges} 
shows the normalized age differences of 56 GCs. The age differences are calculated by subtracting the Paper VII 
age from the age listed in Table \ref{age} and dividing by the quadrature sum of the age uncertainties from both 
sources. In accordance with Paper VII, metallicity is represented in Figure \ref{RelAges} and later figures by 
[M/H]=$\feh$+Log$_{10}(0.638 \times 10^{\afe}+0.362)$ \citep{sa93}. The ages of 43 of the 55 GCs 
shown in Figure \ref{RelAges} differ by less than 1-$\sigma$. The primary reason for disagreement in ages is 
most likely the adopted metallicity scale since the largest systematic deviation occurs around [M/H]$\sim -1$ 
and the metallicity scales deviate there by as much as 0.3 dex (see Figure \ref{met}).

\begin{figure}
\epsscale{0.9}
\plotone{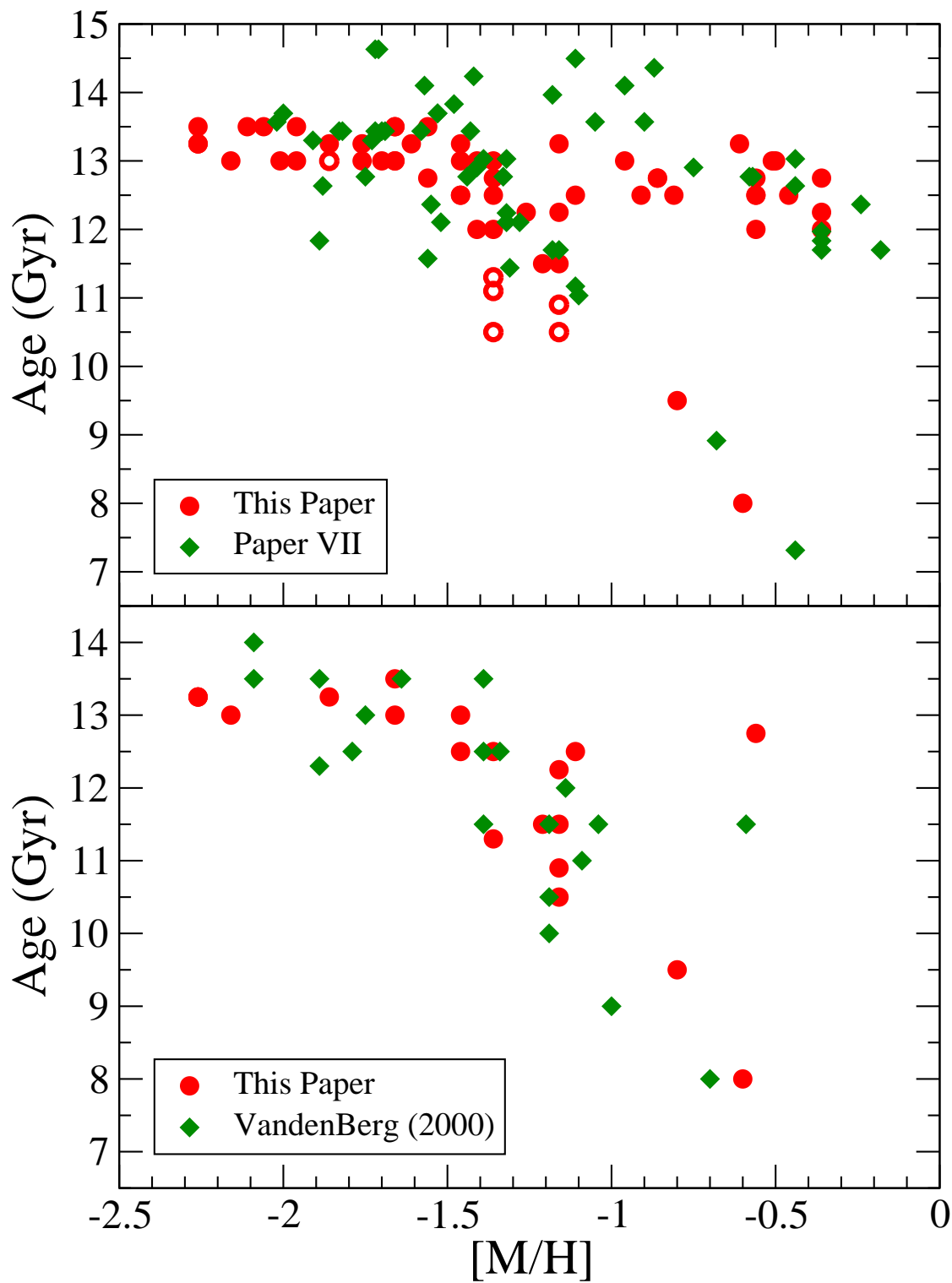}
\caption{{\it Top:} Age-metallicity relations from the present study and Paper VII for the 55 GCs 
in common. The open circles are the outer halo GCs. {\it Bottom:} Age-metallicity relations
from the present study and \citet{va00} for the 20 GCs in common.\label{AMR}}
\end{figure}

Figure \ref{AMR} compares the AMRs from this paper with those of Paper VII (top panel) and 
\citet{va00} (bottom panel). Despite the age differences already addressed in Figure \ref{RelAges} 
and the preceding paragraph, both AMRs reveal the same general features.  The main difference is that
the ages from Table \ref{age} show smaller dispersion at low and high metallicities. The appearance
of a separate trend beginning at [M/H]=$-1.5$ and extending to Pal~12 and Ter~7 (the two youngest GCs)
is nearly identical in both AMRs.  The open circles in the upper panel are the 
outer halo GCs not present in the ACS Survey data; these additional GCs strengthen the trend
already present in the ACS Survey data. Given that \citet{va00} carried out his analysis 
using a heterogeneous collection of data in $B-V$ and $V-I$, and used distances set by 
the level of the HB, it is encouraging that all three AMRs reveal the same basic trends.

\section{Results \& Analysis\label{results}}

\subsection{$\dvi$ and the second parameter\label{2ndP}}

\begin{figure}
\plotone{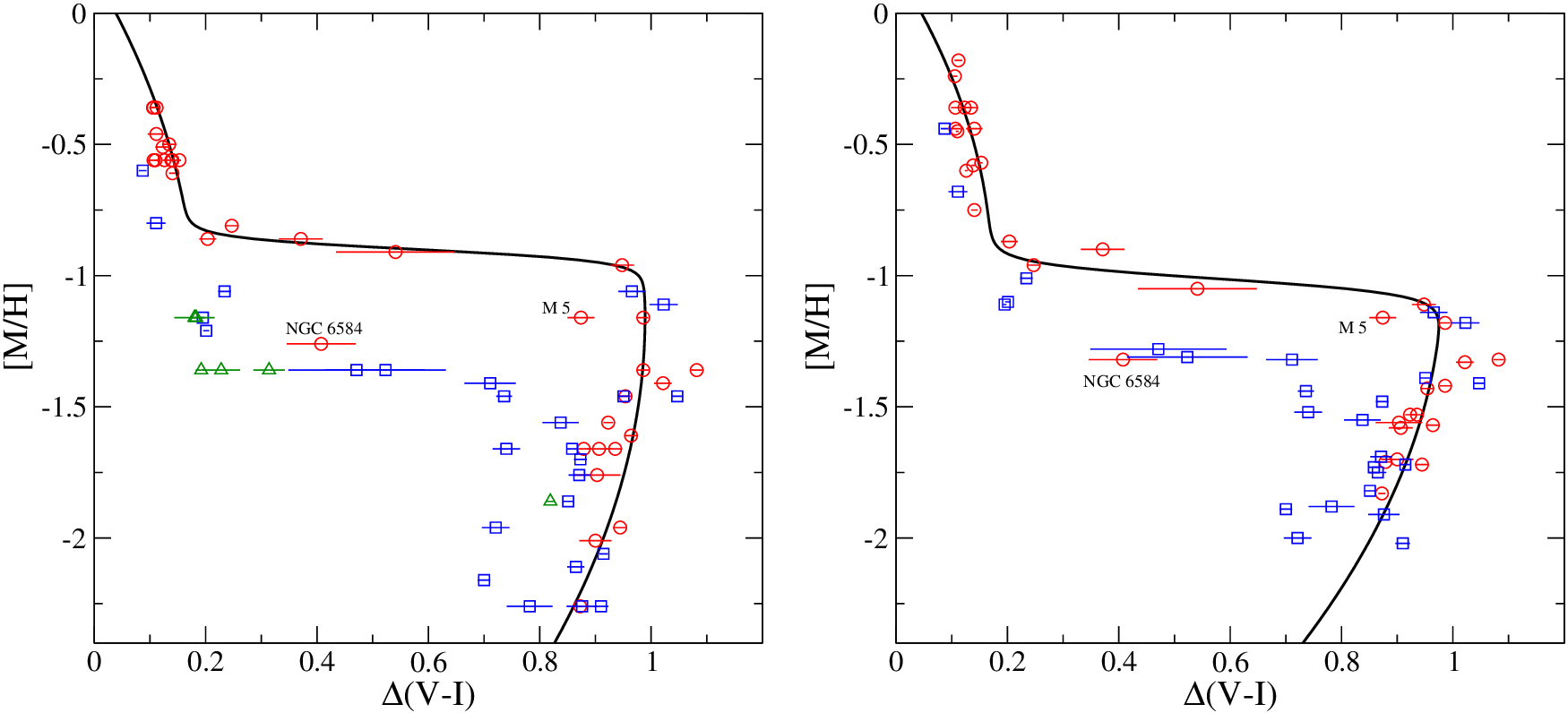}
\caption{The metallicity-HB morphology diagram showing the inner (circles) and outer (squares) halo GCs
from the ACS Survey.  The six most distant GCs are shown as triangles. Error bars are from $\S$\ref{deltavi}; 
fitting functions are shown as solid lines. The left panel shows the metallicity scale from Table \ref{age}; 
the right panels shows the ZW84 metallicity scale from Paper VII.\label{dVIMH}}
\end{figure}

HB morphology is most strongly influenced by the first parameter, metallicity.  The metallicity-$\dvi$ 
diagram is shown in Figure \ref{dVIMH}. For the ACS Survey data, the inner halo GCs ($\rgc \le 8$ kpc) are 
plotted as circles and the outer halo GCs ($\rgc > 8$ kpc) as squares.  The six most distant GCs are plotted 
as triangles and only appear in the left panel because they were not considered in Paper VII.  The left panel 
shows the metallicity scale from Table \ref{age} and the right panel shows the ZW84 metallicity scale from Paper 
VII. The error bars shown are only due to measurement errors in $\dvi$ as listed in Table \ref{morph}. 

In order to more clearly identify the second parameter, an attempt was made to remove the influence of 
metallicity from the HB morphology data, thereby exposing the second parameter effect. To accomplish this, 
a function was fit to the inner halo GCs which cover the full range of metallicity and exhibit relatively 
little scatter in the [M/H]-$\dvi$ diagram, with the notable exception of NGC~6584.  NGC~6584 is presently 
located at $\rgc \sim 7$ kpc but \citet{ldz94} noted its position in the metallicity-HB morphology diagram 
and large, positive heliocentric velocity and suggested it is actually an outer halo GC. This suggestion is 
supported by \citet{di99} who calculated its apogalacticon at $12.6\pm2.4$ kpc. As a result, M~5 (see 
$\S$\ref{intro}) and NGC~6584 were excluded from the fit.  The fitting function is made up of two parts with
$x$=[M/H],
\begin{equation}\label{fit0}
\dvi_{fit}(x) = f(x) + g(x).
\end{equation}
The first part, $f$, resembles a Fermi-Dirac function that transitions from blue to red as 
metallicity increases, as suggested by \citet{ca05}. The second part, $g$, is a quadratic that allows for the 
remaining curvature, in particular that the most metal-poor GCs turn back to the red with decreasing 
metallicity as can be seen in Figure \ref{dVIMH}. The two parts are 
\begin{equation}\label{fit1}
f(x) = a_0 - a_1 \left[\frac{exp\left(\frac{x+a_2}{a_3}\right)}{1 + exp\left(\frac{x+a_2}{a_3}\right)}\right]
\end{equation}
and
\begin{equation}\label{fit2}
g(x) = b_0 + b_1~x + b_2~x^2.
\end{equation}
The $a$'s and $b$'s were determined from a least squares fit. 
For comparison, and to demonstrate the robustness of the method, the same functional
form was fit to the inner halo $\dvi$ data using the metallicity scale from Table \ref{age} (shown
in the left panel of Figure \ref{dVIMH}) and the ZW84 metallicity scale from Paper VII (right panel).
The fitting function coefficients are reproduced in Table \ref{coeff}.

\begin{figure}
\plotone{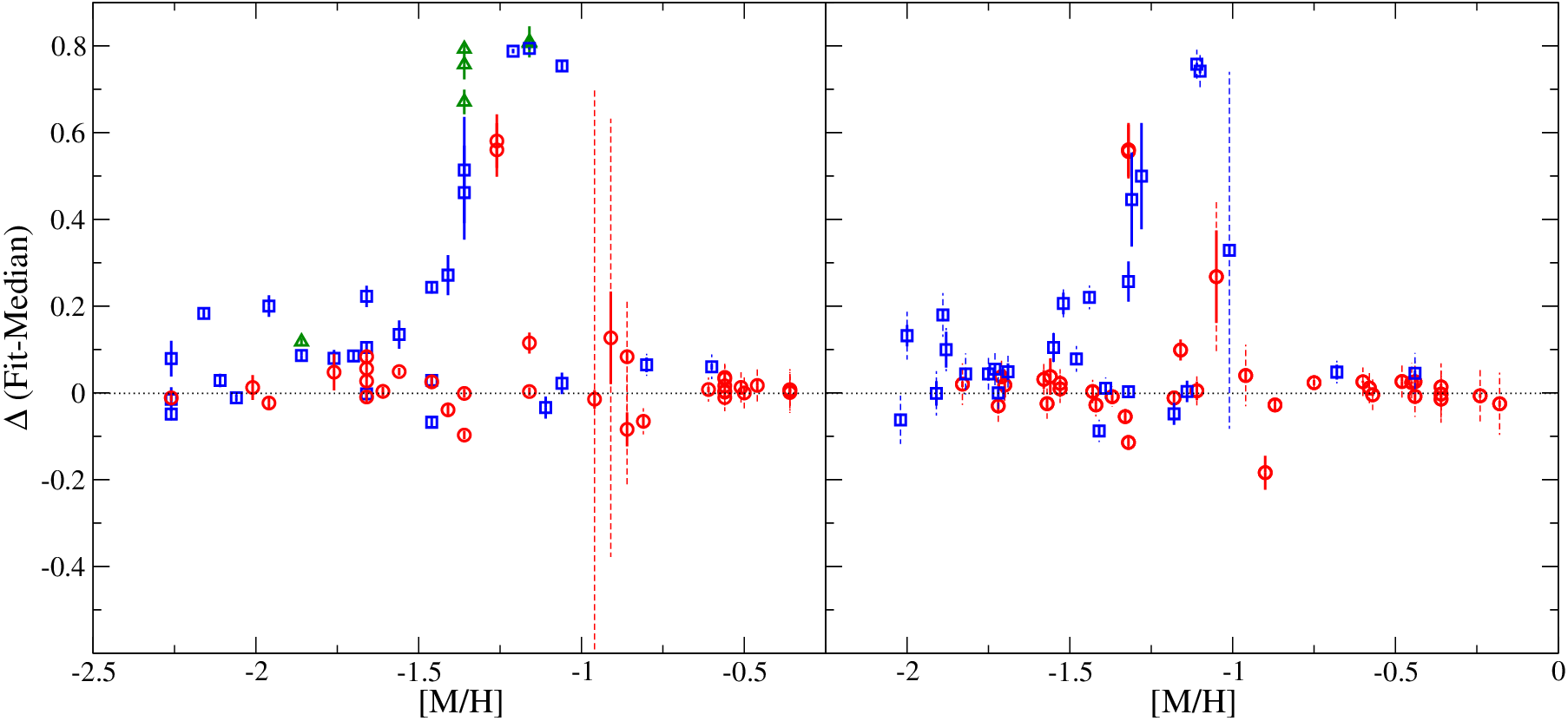}
\caption{Difference between fitted and measured $\dvi$ values as a function of [M/H].
The symbols and layout are the same as in Figure \ref{dVIMH}. The solid error bars are from measurement
uncertainty in $\dvi$ alone; the dashed error bars add the effect of metallicity uncertainty.
\label{dHBMH}}
\end{figure}

The search for correlations focuses on the difference between the measured $\dvi$ of an outer halo GC 
and the $\dvi$ value of the inner halo trend (as represented by the fitting function) at that GC's metallicity.
This quantity will henceforth be referred to as $\dfm$. 
Figure \ref{dHBMH} shows how $\dfm$ varies with metallicity; the symbols are the same as in Figure \ref{dVIMH}.
With the exception of NGC~6584 the inner halo GCs exhibit residuals of $\sim$0.1 or less. Altogether, metal-rich 
GCs ([M/H]$\ga-0.8$) show very little scatter in Figure \ref{dHBMH}. Evidently, metallicity alone is almost 
sufficient to describe the HB morphology of the majority of metal-rich GCs and thus they are of little 
use in the search for the second parameter. The solid error bars shown in Figure \ref{dHBMH} represent the 
measurement errors listed in Table \ref{morph}.

Metallicity uncertainty plays an important role in this diagram and thus the error bars in Figure \ref{dHBMH} 
 are only a lower limit to the total uncertainty in $\dvi$.
Unfortunately, it is not possible to provide realistic metallicity errors for all of the GCs in 
the sample. For the large deviation shown for $-1.5 <$ [M/H] $< -1$ in Figure \ref{dHBMH} to be completely 
erased, metallicities of the GCs in this region that stand out would have to be systematically under-estimated 
by 0.25-0.5 dex. Such a conspiracy seems highly unlikely. However, it is possible to approximate the effect
of metallicity errors on $\dvi$ in the following manner. First, the fitting function described by equations
\ref{fit1} and \ref{fit2} was applied to the inner halo GCs with metallicities on the KI03 scale.
Next, the KI03-based fitting function was used to calculate a $\dvi$ but using the metallicity 
from Table \ref{age} or the ZW84 scale, thereby introducing a metallicity error. Finally, the difference between 
the erroneous $\dvi$ value obtained in this manner and the value obtained from the appropriate fitting function 
provides an estimate of how a metallicity error propagates into a $\dvi$ error.

Under the hypothesis that the KI03 $\feh$ scale is the `correct' one, this procedure gives an indication of how a 
metallicity error (a vertical displacement in Figure \ref{dVIMH}) influences the significance of the distance 
between a GC and the inner halo fitting function (the horizontal distance between a point and the fit line in 
Figure \ref{dVIMH}) which we refer to as $\dfm$. The dashed error bars in Figure \ref{dHBMH} are the quadrature 
sum of the $\dvi$ measurement errors from Table \ref{morph} and the metallicity errors estimated as described 
in the preceding paragraph. The metallicity effect is small except in the vicinity of the transition from red 
to blue that occurs around [M/H] $\sim -1$ (see Figure \ref{dVIMH}) where a small change in metallicity corresponds 
to a large change in $\dvi$. It should be stressed that this approach is not a rigorous treatment of the influence 
of metallicity uncertainty on $\dvi$ or $\dfm$ because it assumes the difference between one scale and another 
is the error.  Nevertheless, it serves to illustrate that the existence of several GCs with $-1.5 \la$ [M/H] 
$\la -1$ and large, positive values of $\dfm$ cannot readily be attributed to metallicity errors alone.

\begin{figure}
\plotone{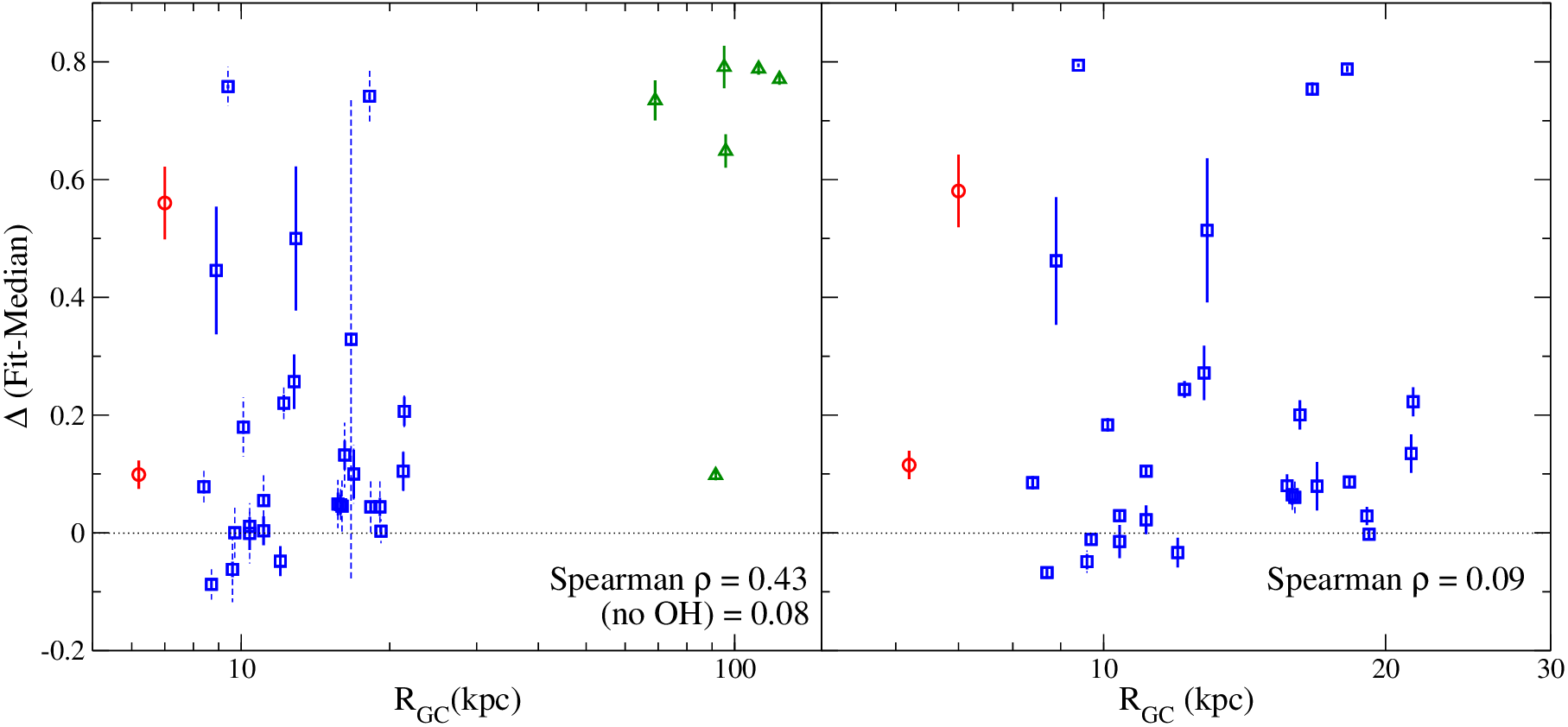}
\caption{The difference between fit and measured $\dvi$ values as a function of $\rgc$. Symbols and 
layout are as in Figure \ref{dHBMH}. The circles are M~5 (near the dotted line) and NGC~6584 (further away). 
Included on the figure is the value of the Spearman rank coefficient, see text for details. The phrase `no OH' 
means that the 6 most distant GCs were omitted from the Spearman coefficient calculation.\label{dHBRgc}}
\end{figure}

\begin{figure}
\plotone{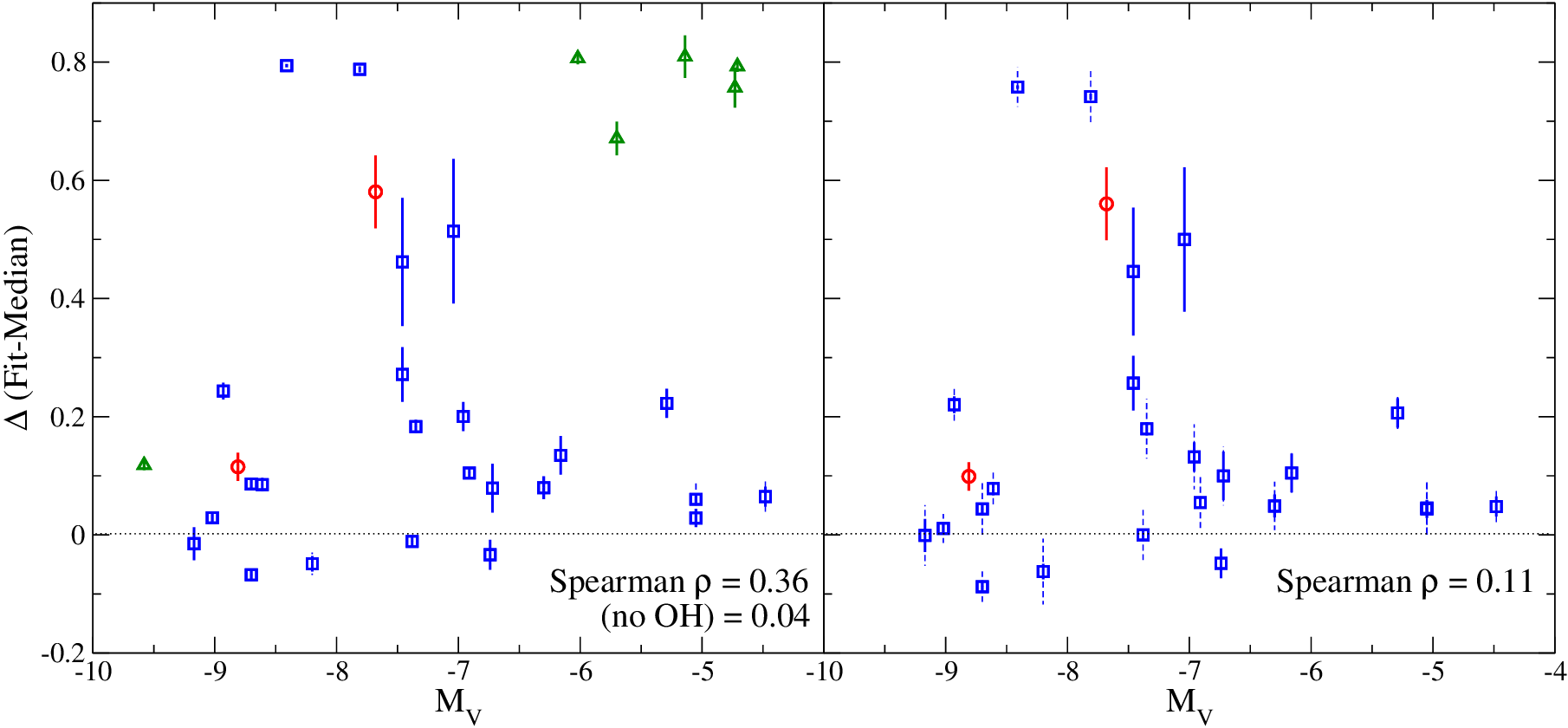}
\caption{Similar to Figure \ref{dHBRgc} but as a function of $\mv$.\label{dHBMv}}
\end{figure}

\begin{figure}
\plotone{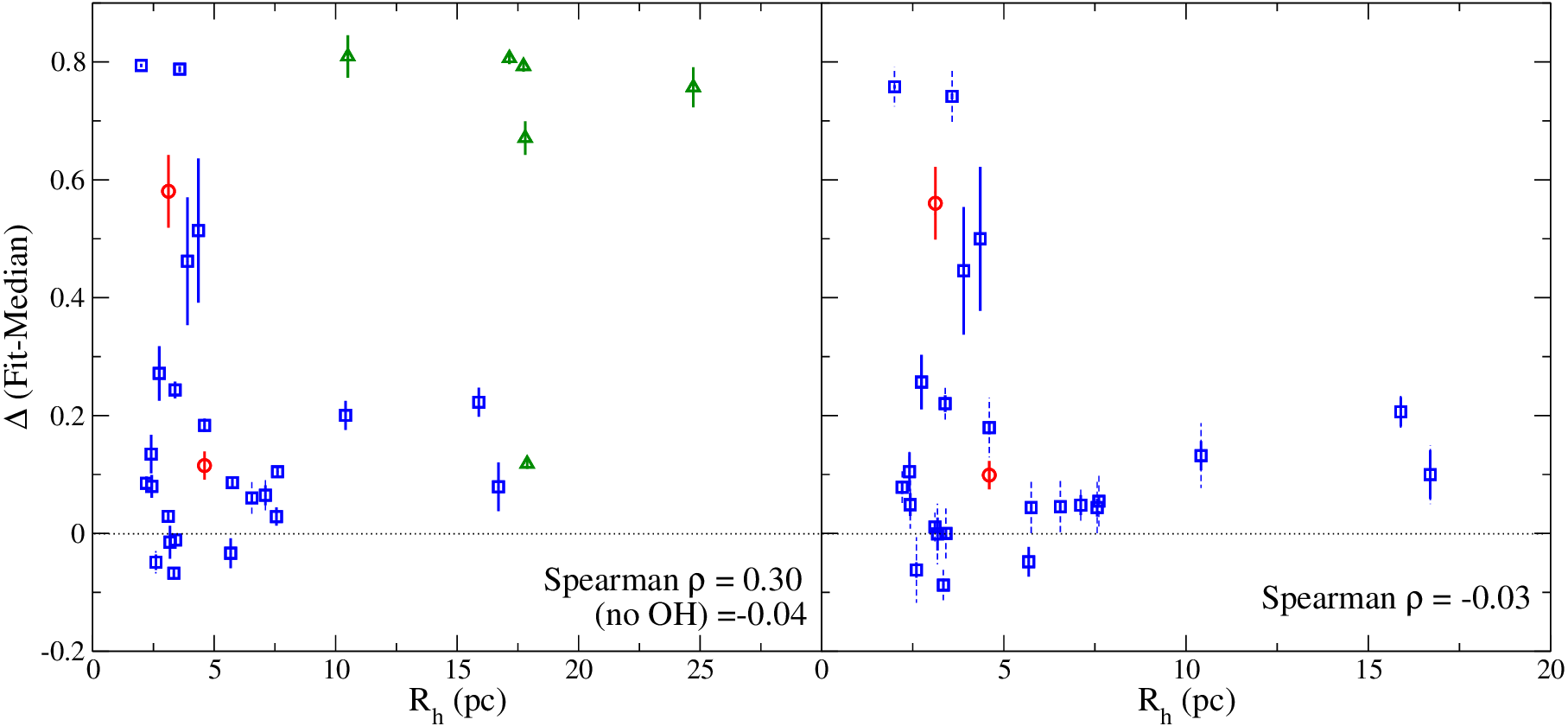}
\caption{Similar to Figure \ref{dHBRgc} but as a function of $\rh$.\label{dHBRh}}
\end{figure}

\begin{figure}
\plotone{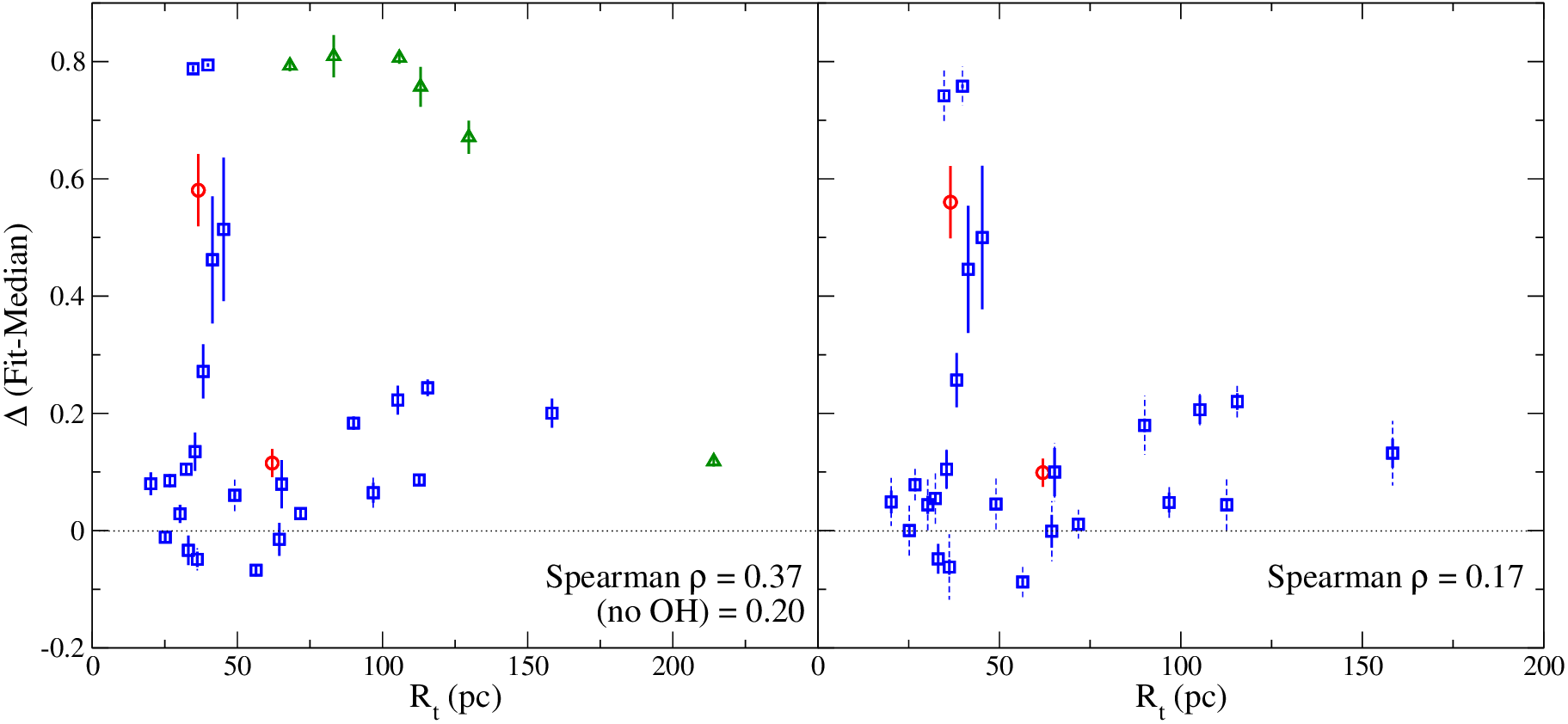}
\caption{Similar to Figure \ref{dHBRgc} but as a function of $\rt$.\label{dHBRt}}
\end{figure}

\begin{figure}
\plotone{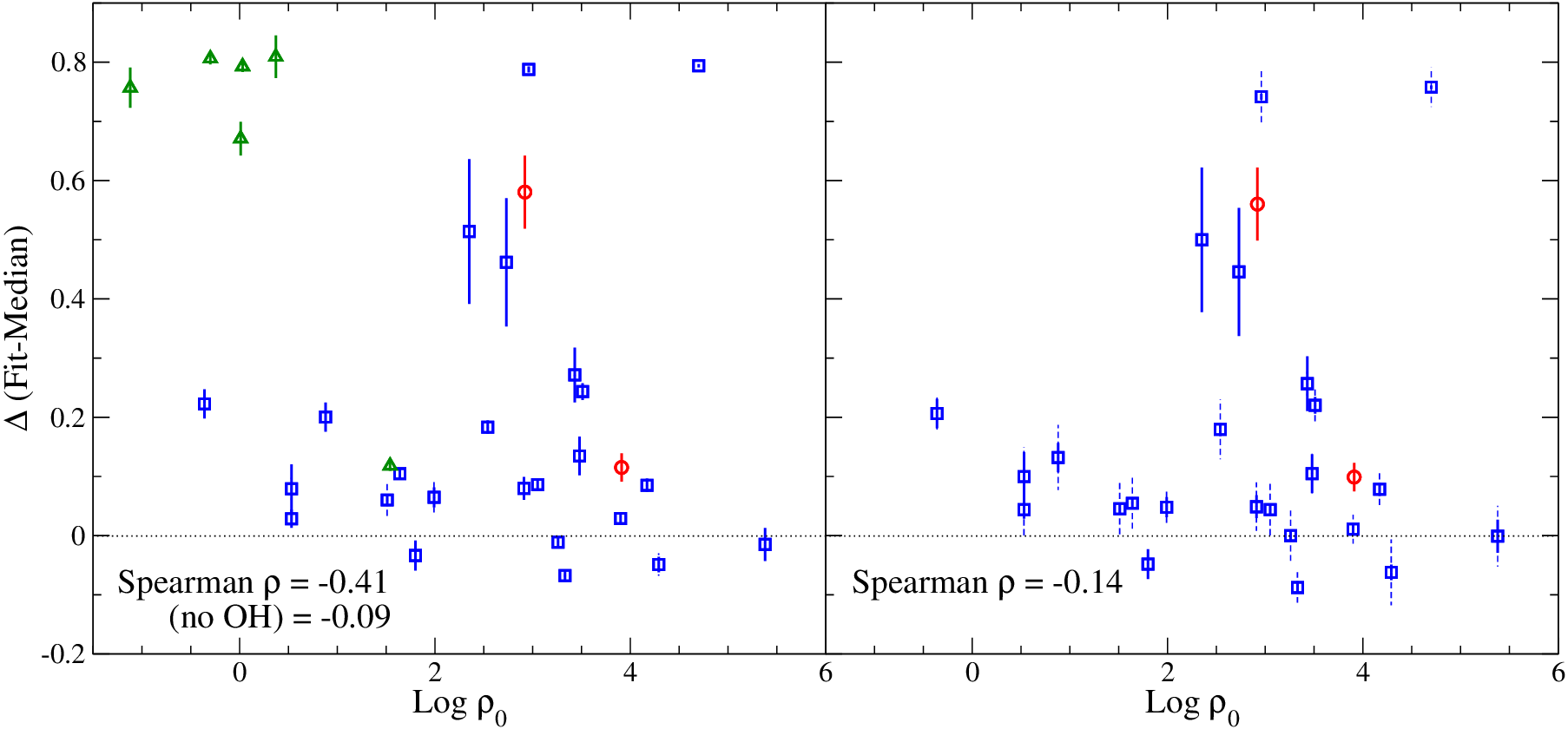}
\caption{Similar to Figure \ref{dHBRgc} but as a function of Log~$\rho_0$.\label{dHBrho}}
\end{figure}

\begin{figure}
\plotone{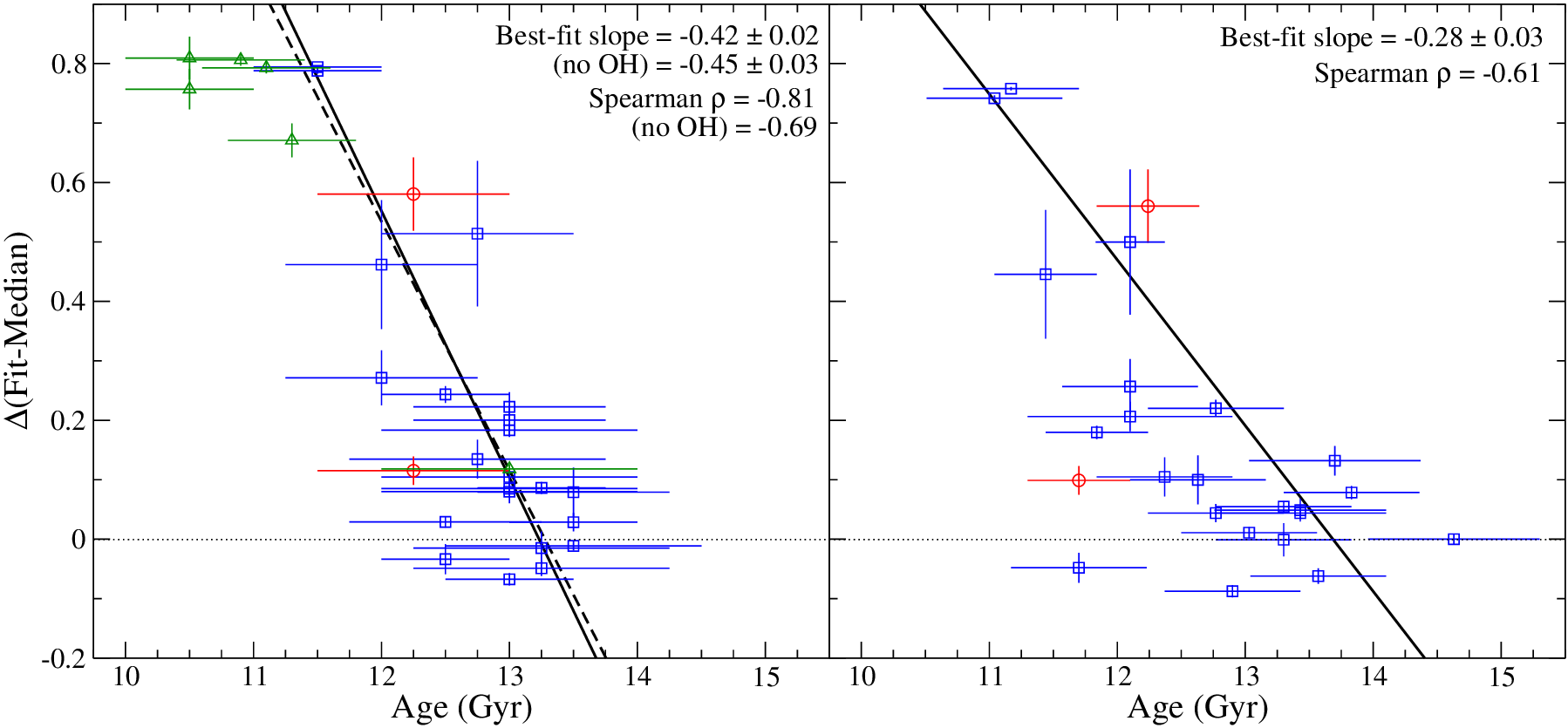}
\caption{Similar to Figure \ref{dHBRgc} but as a function of age. The left panel shows the 
ages reported in Table \ref{age} while the right panel shows the D07/ZW84 ages from Paper VII. 
The solid line is a fit to the ACS Survey data alone; the dashed line is a fit that includes
the six outer halo GCs. The two youngest GCs, Pal~12 and Ter~7, have been excluded from this 
figure and calculation of the coefficients, see text for discussion.\label{dHBAge}}
\end{figure}

Figures \ref{dHBRgc} through \ref{dHBAge} show how the $\dvi$ difference depends on 
$\rgc$, $\mv$, $\rh$, $\rt$, $\rho_0$, and age for the outer halo GCs. In order to quantify the information
provided in the plots, Spearman's rank correlation coefficient was calculated for the data presented in 
Figures \ref{dHBRgc} through \ref{dHBAge} and included in the figures.
The Spearman rank correlation coefficient (Spearman's $\rho$) measures the degree of correlation
between two variables but makes no assumption about the functional form of their relationship other than
monotonicity.  As such, it is a very general measure of correlation; a perfect correlation has a value of
$\rho = +1$, a perfect anti-correlation has $\rho = -1$, and complete lack of correlation has $\rho = 0$.
The left panels show the 6 most distant GCs, which were not part of the ACS Survey, as triangles. 
Since these 6 GCs are not part of the homogeneous ACS data set, and were not considered in Paper VII, we have 
calculated the Spearman correlation coefficients with and without them in the left panels.  
Apart from NGC~2419, the other 5 most distant GCs tend to strengthen the correlation coefficients because they 
are more distant, intrinsically fainter, more extended, and younger.

$\dvi$ error bars are included in Figures \ref{dVIMH} through \ref{dHBAge}, and age errors in Figure 
\ref{dHBAge}, but uncertainties in the other quantities are neither readily available nor simple to estimate
given the wide range of distances and reddenings present in the sample. It is, however, important to consider 
that the quantities in Figures \ref{dHBRgc} through \ref{dHBrho} were converted to physical units
after adopting a particular distance modulus. As distance increases, a given error in the distance 
corresponds to a larger uncertainty in the physical quantity.

While $\rgc$, $\mv$, $\rh$, $\rt$, and $\rho_0$ reveal no obvious visible trends, Figure \ref{dHBAge} shows a
trend with age. The Spearman coefficients support this conclusion: $\mv$, $\rh$, $\rt$, and $\rho_0$ show no 
strong correlations. $\rgc$ shows a mild correlation but that is mostly likely due to a large anti-correlation
between age and $\rgc$. Age shows a significant anti-correlation and the age trend works as anticipated: 
the HBs grow redder with decreasing age and so the distances between the measured $\dvi$ values and the inner 
halo trend increase as age decreases.
The results presented in this section demonstrate that, of all the parameters considered here, age has the most 
significant correlation with $\dfm$.

Figure \ref{dHBAge} and the accompanying slopes and coefficients do not include Pal~12 or Terzan~7, the two 
youngest GCs in the sample.  Both GCs are metal rich and so their HB morphologies are almost entirely determined 
by their metallicity alone. The presence or absence of Pal~12 and Ter~7 does not significantly alter the 
correlation coefficients ($\delta \rho \la 0.05$) for any of the other quantities. Figure \ref{dHBAge} also 
includes a linear fit to each data set that describes the rate at which HB morphology varies with age.  The ages 
and metallicities listed in Table \ref{age} suggest that HB morphology changes over a smaller age interval than 
those from Paper VII, but only by about $\sim$0.5 Gyr.  The larger errors that include
the metallicity effect were not used in the least squares fits to the age-$\dfm$ relations shown in Figure 
\ref{dHBAge}.  If they had been, the slope of the relation derived using the Paper VII ages (shown in the right 
panel) would be reduced to $-0.21\pm0.05$ while the slope derived from the ages given in Table \ref{age} would not 
change significantly.

\subsection{A possible third parameter\label{3rdP}}

In $\S$\ref{2ndP}, it was shown that when the inner halo relationship between metallicity and $\dvi$
is subtracted from the outer halo GCs, a correlation between age and $\dfm$ appears. The inner halo GCs 
were chosen to define the trend because they exhibit a tight relation with little scatter in the HB morphology--metallicity 
diagram \citep[see also Figure \ref{dVIMH} in this paper]{sz78}. Unfortunately, attempts to
use this technique a second time--to subtract off the age trend--were unsuccessful because the remaining 
residuals are generally smaller than the measurement uncertainties in $\dvi$. Thus, to identify a potential 
third parameter, it is necessary to remove the effect of metallicity and age to the fullest extent possible. The 
most metal poor GCs ([M/H] $< -1.5$) in the sample are an ideal choice because they have a small range in ages 
($\S$\ref{gcages}) and only a weak metallicity dependence on HB morphology (see Figure \ref{dVIMH}).

\begin{figure}
\plotone{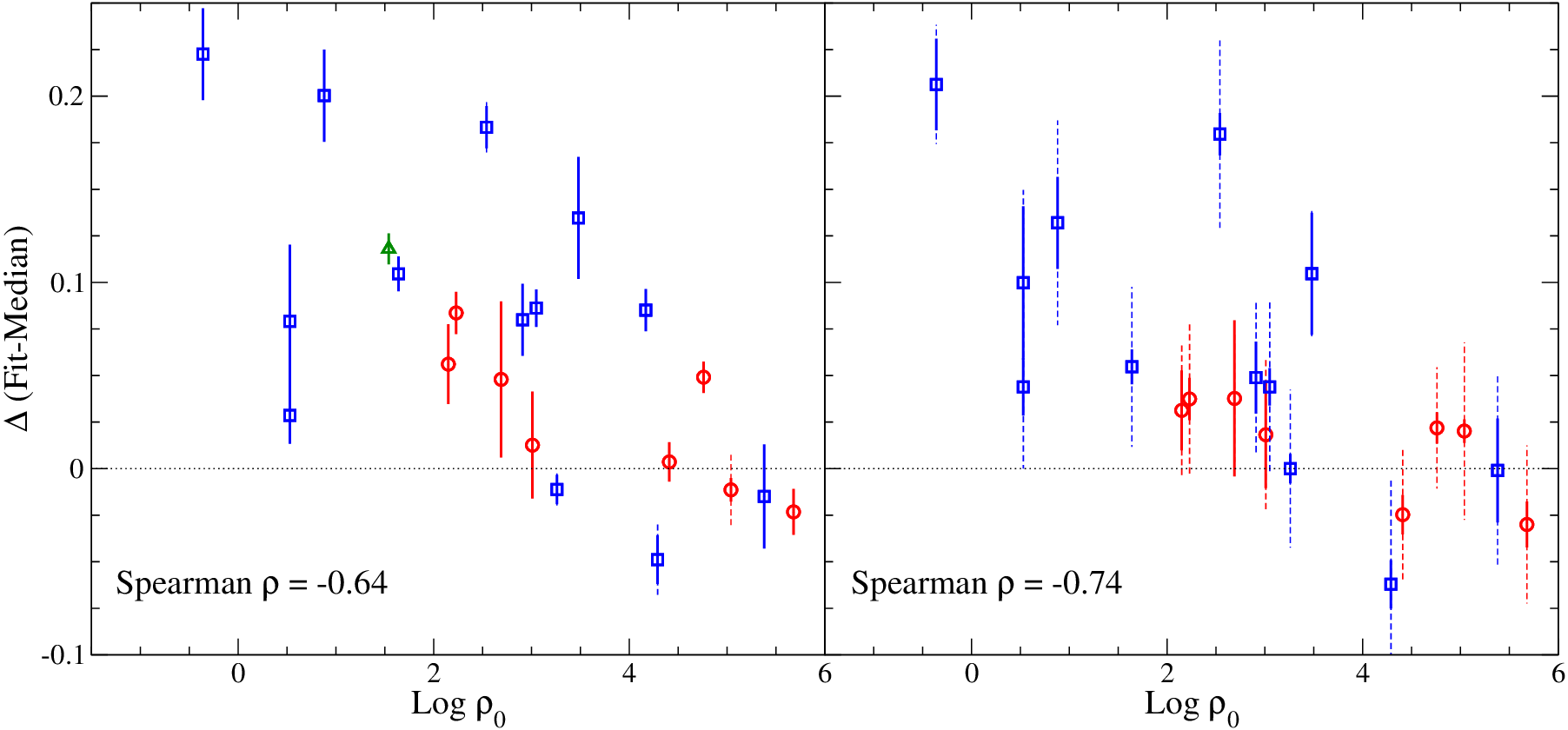}
\caption{Central density (Log~$\rho_0$) vs. $\dfm$ showing only the metal poor GCs 
([M/H] $< -1.5$).  As with Figures \ref{dHBMH} through \ref{dHBAge}, the left and right panels
show results based on the two fitting functions determined in $\S$\ref{2ndP}.
Symbols are the same as in Figure \ref{dVIMH}.\label{rho}}
\end{figure}

Comparisons of the metal poor GCs' $\dfm$ values with $\mv$, $\rh$, $\rt$, and $\rho_0$
were performed. Of these, central density (Log~$\rho_0$) produced the most significant trend. 
Figure \ref{rho} demonstrates how central density relates to $\dfm$ 
among the metal poor GCs using the two fitting functions employed in $\S$\ref{2ndP} and the same symbols 
used in Figures \ref{dVIMH} through \ref{dHBAge}. There is an evident trend with the highest central 
density GCs having the bluest HB morphologies and hence the smallest $\dfm$ values. Assuming that the 
contributions of age and metallicity are minimal in Figure \ref{rho}, the influence that central density 
has on HB morphology in the metal poor GCs is evident and its magnitude is $\sim0.2$ in $\dfm$. Although 
it is difficult to disentangle the effects of age and central density in GCs with intermediate 
metallicities ($-1.5 <$ [M/H] $< -1$), where the second parameter is most pronounced, these GCs span 
essentially the same range of central densities as the metal poor GCs.  Thus central density should only 
account for a small portion of the HB morphology variation seen in the intermediate metallicity group.
The influence of central density on HB morphology has already been demonstrated by \citet{fp93}, 
\citet{bu97}, and \citet{sm04}. However, its effect is not as pronounced as that of age (compare Figures 
\ref{dHBAge} and \ref{rho}). Therefore central density is the most likely candidate for the third 
parameter influencing HB morphology (as characterized by $\dvi$) that has been considered by this study.

\subsection{Discussion and comparisons with previous results}

\begin{figure}
\plotone{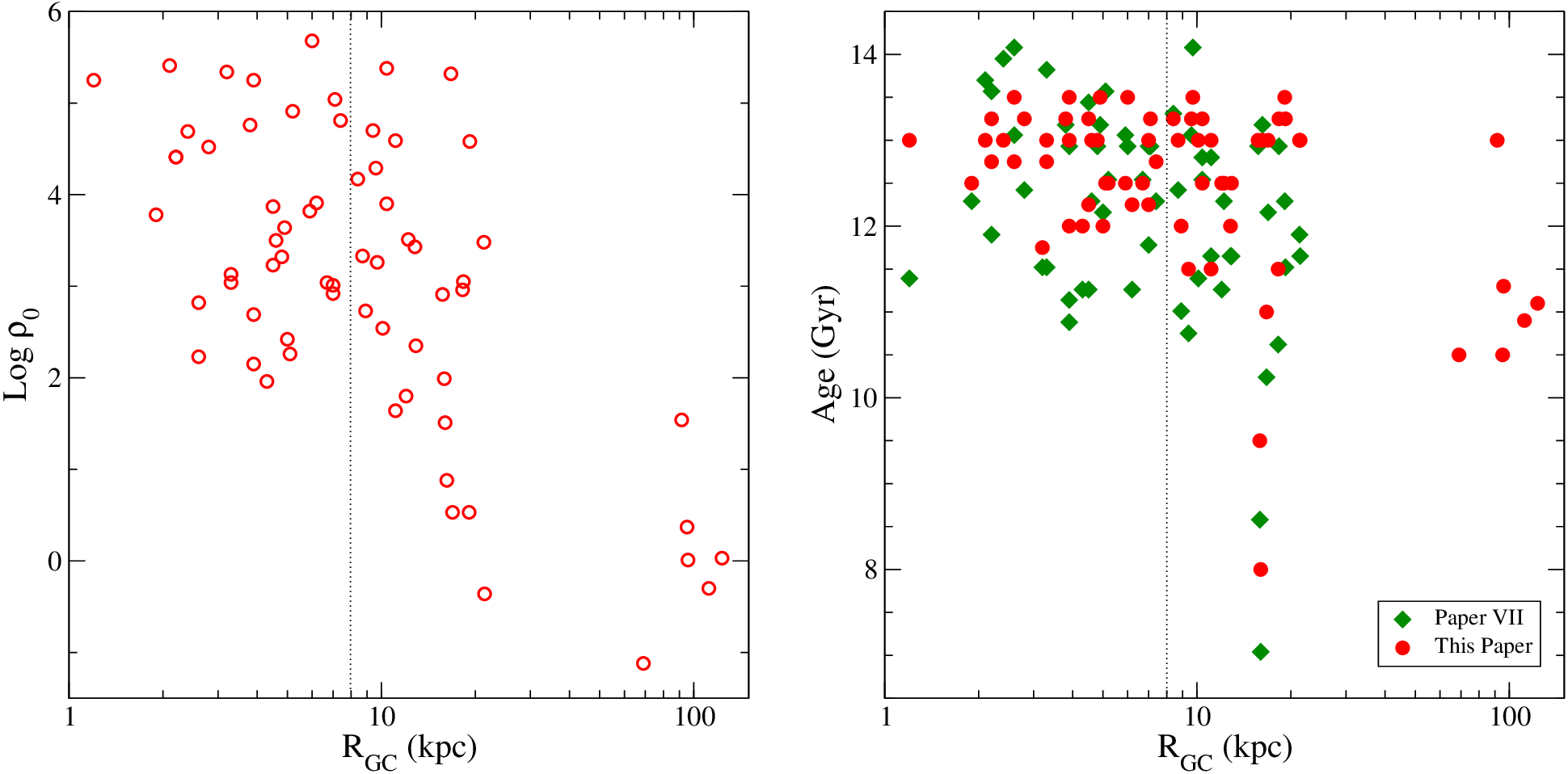}
\caption{The dependence of central density (left panel) and age (right panel) on $\rgc$. The boundary 
assigned between the inner and outer halo is indicated by the dotted line at 8 kpc.  The age plot shows
results from both $\S$\ref{isoage} (circles) and Paper VII (diamonds). $\rgc$ and $\rho_0$ data are from
the Harris catalog.\label{inner}}
\end{figure}

It is worthwhile to demonstrate that the use of the inner halo GCs to define the HB morphology--metallicity 
relation does not include an unintended bias.  In particular, since claims have been made in the preceding 
sections that, after metallicity, age and central density are the two most significant factors influencing 
HB morphology [as represented by $\dvi$], it is important to demonstrate that the HB morphology--metallicity 
trend derived from the inner halo GCs does not include an implicit dependence on either of these quantities. 
Indeed, as Figure \ref{inner} shows, the inner halo GCs exhibit some dispersion inside 8 kpc in both central 
density (left panel) and age (right panel) but there is no clear, systematic variation of either parameter as 
a function of $\rgc$ inside 8 kpc (denoted by the dotted line in the figure).

Figure \ref{dVIMH} demonstrates that the HB morphologies of the inner halo GCs, which have a fairly 
homogeneous age distribution, transition from blue to red as [M/H] rises through $\sim -1$. The metal 
rich GCs have red HBs (though NGC~6388 and NGC~6441 also have blue extensions) while the majority of
metal poor GCs have blue HBs. 
It is at metallicities that are lower than, but within a few tenths of a dex of, [M/H]=$-1$ where a GC 
is most likely to move a large distance in the $\dvi$--metallicity diagram due to a relatively small change in
metallicity or another parameter. It is within this same narrow metallicity range that 
the AMR of the outer halo GCs branches off from that of the inner halo GCs (see Figure \ref{AMR}) 
and so it is no coincidence that the classic second parameter effect is most pronounced in the outer halo.

The analysis presented in $\S$\ref{2ndP} indicates that a GC with $-1.5 <$[M/H]$< -1$ requires 
2-2.5 Gyr to transition from a red HB to a blue one. This result is consistent with a number 
of studies mentioned in the introduction.  We confirm the assertion by \citet{ldz94} that the presence 
of metal poor GCs with red HBs at large $\rgc$ is primarily an age effect. However, our result 
indicates that Lee et al.'s claim that the necessary age difference is greater than $\sim$3 Gyr
is slightly exaggerated\footnote{It is important to keep in mind that there is a small, but
not insignificant, number of GCs with younger ages, 
such as Pal~1 (Paper I), Pal~12, and Ter~7; these tend to be metal rich and so their HB morphologies
are governed almost entirely by metallicity.}. That a GC with $-1.5 <$ [M/H] $< -1$ transitions from 
a red to a blue HB in 2-2.5 Gyr poses a challenge to our understanding of mass loss during the red giant 
phase of evolution. RGB mass loss has been essentially a free parameter in
synthetic HB models since they were first constructed by \citet{ro73}. A number
of mass loss rates that vary as a function of global stellar properties such as luminosity, mass, radius,
or some combination of the three have been proposed with varying degrees of success, see \citet{ca05} for
a detailed discussion. Following \citet{le91}, \citet{do08c} showed that a simple relationship between 
global metallicity and average RGB mass loss, along with stochastic variations, can reproduce the general
trend observed in old, roughly coeval Galactic GCs in the HB morphology-metallicity diagram.
Recent progress from observations by, e.g.\,\citet{or07}, \citet{mc09}, \citet{me09}, and \citet{du09},
should lead to a better understanding of RGB mass loss as a function of composition, evolutionary status, 
and pulsational properties.

The recent, large-scale study of HB morphology by \citet{rb06}  suggests that, just as age can explain anomalously 
red HBs in some GCs, total GC mass is linked to the degree of abundance variation and the extent of the faint blue 
tail observed in other GCs. (Of course, there is no reason why both effects cannot be operating in the same GC.)
Recall that Figure \ref{dVIBVR} showed the HB morphology metric adopted by \citet{rb06} was very sensitive to the 
extremes but lacked sensitivity through the middle of the distribution. In the absence of an ideal HB morphology 
metric, a large-scale study of HB morphology should choose a metric that is well-suited to the effect(s) of interest 
to that investigation.

\section{Conclusion\label{conclusion}}
HB morphologies characterized by $\dvi$ of 66 Galactic GCs were examined to determine
the sensitivity of HB morphology to a variety of different factors. Deep, homogeneous photometry from
the ACS Survey of Galactic GCs accounts for 60 of these while the remaining 6 GCs are the most distant
Galactic GCs known. The complete sample is the largest examined to date and consists solely of high-quality 
HST photometry. It spans the full range of $\rgc$ and almost the entire range of metallicity present in the 
Galactic GC population. $\dvi$ values and two independently measured sets of ages were joined with other 
quantities from the literature to assess the relative importance of these quantities on HB morphology. 

The data were split into two groups, roughly equal in number, based on $\rgc$. The tight relationship 
between metallicity and HB morphology in the inner halo group ($\rgc < 8$ kpc) was characterized by 
a fitting function and this trend was subtracted off of the outer halo group.
The difference between fit and measured $\dvi$ was then compared to a variety of parameters, of which
only age showed a significant correlation. The age correlation does not rely on the presence of the 6 
most distant GCs in the analysis, though their presence does strengthen the result.  Hence we conclude that, 
after metallicity, age has the most influence on $\dvi$.  The age spread among the bulk of GCs in the Galactic 
halo was found to be 2-2.5 Gyr, though there are a few younger outliers such as Pal~12 and Ter~7. 
Further analysis, in which both metallicity and age were restricted, provided strong evidence that central 
luminosity density ($\rho_0$) is the third most influential parameter on $\dvi$.

\acknowledgments
We wish to express our gratitude to the anonymous referee for a thoughtful, thorough report that improved the 
accuracy and focus of the paper. We thank Ivan King for sharing his differential reddening corrected CMDs in 
advance of publication and the following people for sharing their HST photometry: Peter Stetson (Pal~3, Pal~4, 
and Eridanus), Bill Harris (NGC~2419), and Eric Sandquist (NGC~2419). 
AD acknowledges support from a CITA National Fellowship and an NSERC grant to D. VandenBerg.  Further support 
for this work (proposal number GO-10775) was provided by NASA through a grant from the Space Telescope Science 
Institute which is operated by the Association of Universities for Research in Astronomy, Incorporated, under 
NASA contract NAS5-26555.

\clearpage

\clearpage

\clearpage

\clearpage

\clearpage
\clearpage
\clearpage
\clearpage
\clearpage
\clearpage

\clearpage
\begin{deluxetable}{cccccc}
\tabletypesize{\footnotesize}
\tablecolumns{6}
\tablewidth{0pc}
\tablecaption{Horizontal Branch Parameters\label{morph}}
\tablehead{\colhead{}&\colhead{}&\colhead{$(V-I)_{\mathrm{HB}}$}&\colhead{$(V-I)_{\mathrm{HB}}$}&
\colhead{$(V-I)_{\mathrm{RGB}}$}&\colhead{}\\
\colhead{ID}&\colhead{F606W(HB)}&\colhead{Median}&\colhead{MAD}&\colhead{Median}&\colhead{$\dvi$}}
\startdata
Arp~2    & 18.05 &  0.316 $\pm$  0.017 &  0.200 &  1.056 $\pm$  0.018 &  0.740 $\pm$  0.025 \\
Lyng\aa7 & 17.10 &  1.863 $\pm$  0.003 &  0.034 &  1.974 $\pm$  0.015 &  0.111 $\pm$  0.015 \\
NGC~104  & 13.90 &  0.916 $\pm$  0.002 &  0.023 &  1.069 $\pm$  0.002 &  0.153 $\pm$  0.003 \\
NGC~288  & 15.35 & -0.018 $\pm$  0.023 &  0.127 &  1.004 $\pm$  0.011 &  1.022 $\pm$  0.025 \\
NGC~362  & 15.33 &  0.786 $\pm$  0.003 &  0.083 &  0.981 $\pm$  0.001 &  0.195 $\pm$  0.003 \\
NGC~1261 & 16.60 &  0.762 $\pm$  0.004 &  0.129 &  0.963 $\pm$  0.003 &  0.201 $\pm$  0.005 \\
NGC~1851 & 16.00 &  0.764 $\pm$  0.011 &  0.272 &  0.998 $\pm$  0.002 &  0.234 $\pm$  0.011 \\
NGC~2298 & 16.00 &  0.332 $\pm$  0.018 &  0.123 &  1.203 $\pm$  0.007 &  0.871 $\pm$  0.019 \\
NGC~2808 & 16.10 &  0.235 $\pm$  0.025 &  0.439 &  1.201 $\pm$  0.002 &  0.966 $\pm$  0.025 \\
NGC~3201 & 14.60 &  0.767 $\pm$  0.108 &  0.270 &  1.290 $\pm$  0.011 &  0.523 $\pm$  0.108 \\
NGC~4147 & 16.85 &  0.103 $\pm$  0.031 &  0.136 &  0.941 $\pm$  0.011 &  0.838 $\pm$  0.033 \\
NGC~4590 & 15.65 &  0.291 $\pm$  0.011 &  0.147 &  0.991 $\pm$  0.003 &  0.700 $\pm$  0.011 \\
NGC~4833 & 15.50 &  0.451 $\pm$  0.029 &  0.154 &  1.351 $\pm$  0.003 &  0.900 $\pm$  0.029 \\
NGC~5024 & 16.80 &  0.090 $\pm$  0.009 &  0.088 &  0.941 $\pm$  0.003 &  0.851 $\pm$  0.010 \\
NGC~5053 & 16.60 &  0.145 $\pm$  0.041 &  0.158 &  0.927 $\pm$  0.003 &  0.782 $\pm$  0.041 \\
NGC~5272 & 15.55 &  0.207 $\pm$  0.014 &  0.208 &  0.943 $\pm$  0.002 &  0.736 $\pm$  0.014 \\
NGC~5286 & 16.40 &  0.372 $\pm$  0.011 &  0.182 &  1.245 $\pm$  0.003 &  0.873 $\pm$  0.011 \\
NGC~5466 & 16.45 &  0.208 $\pm$  0.024 &  0.124 &  0.929 $\pm$  0.006 &  0.721 $\pm$  0.025 \\
NGC~5904 & 14.90 &  0.120 $\pm$  0.024 &  0.230 &  0.994 $\pm$  0.001 &  0.874 $\pm$  0.024 \\
NGC~5927 & 16.35 &  1.490 $\pm$  0.002 &  0.025 &  1.602 $\pm$  0.006 &  0.112 $\pm$  0.007 \\
NGC~5986 & 16.40 &  0.357 $\pm$  0.011 &  0.136 &  1.311 $\pm$  0.004 &  0.954 $\pm$  0.012 \\
NGC~6093 & 16.20 &  0.275 $\pm$  0.008 &  0.153 &  1.198 $\pm$  0.003 &  0.923 $\pm$  0.008 \\
NGC~6101 & 16.50 &  0.203 $\pm$  0.007 &  0.091 &  1.061 $\pm$  0.005 &  0.858 $\pm$  0.009 \\
NGC~6121 & 13.17 &  0.987 $\pm$  0.106 &  0.252 &  1.528 $\pm$  0.010 &  0.541 $\pm$  0.106 \\
NGC~6144 & 16.20 &  0.612 $\pm$  0.009 &  0.100 &  1.491 $\pm$  0.007 &  0.879 $\pm$  0.011 \\
NGC~6171 & 15.40 &  1.296 $\pm$  0.013 &  0.104 &  1.499 $\pm$  0.009 &  0.204 $\pm$  0.015 \\
NGC~6205 & 14.85 & -0.091 $\pm$  0.010 &  0.122 &  0.956 $\pm$  0.002 &  1.047 $\pm$  0.010 \\
NGC~6218 & 14.60 &  0.227 $\pm$  0.009 &  0.099 &  1.213 $\pm$  0.004 &  0.986 $\pm$  0.009 \\
NGC~6254 & 14.80 &  0.269 $\pm$  0.015 &  0.113 &  1.290 $\pm$  0.004 &  1.022 $\pm$  0.016 \\
NGC~6304 & 15.95 &  1.556 $\pm$  0.002 &  0.020 &  1.662 $\pm$  0.004 &  0.105 $\pm$  0.004 \\
NGC~6341 & 15.05 &  0.036 $\pm$  0.013 &  0.093 &  0.946 $\pm$  0.004 &  0.910 $\pm$  0.013 \\
NGC~6352 & 14.95 &  1.249 $\pm$  0.004 &  0.025 &  1.372 $\pm$  0.013 &  0.123 $\pm$  0.013 \\
NGC~6362 & 15.18 &  0.817 $\pm$  0.009 &  0.193 &  1.064 $\pm$  0.007 &  0.247 $\pm$  0.012 \\
NGC~6366 & 15.25 &  1.817 $\pm$  0.010 &  0.030 &  1.957 $\pm$  0.011 &  0.141 $\pm$  0.014 \\
NGC~6388 & 17.00 &  1.344 $\pm$  0.002 &  0.172 &  1.470 $\pm$  0.002 &  0.126 $\pm$  0.003 \\
NGC~6397 & 12.90 &  0.217 $\pm$  0.007 &  0.066 &  1.162 $\pm$  0.010 &  0.944 $\pm$  0.012 \\
NGC~6441 & 17.65 &  1.501 $\pm$  0.002 &  0.154 &  1.611 $\pm$  0.002 &  0.110 $\pm$  0.003 \\
NGC~6496 & 16.20 &  1.224 $\pm$  0.004 &  0.023 &  1.331 $\pm$  0.007 &  0.107 $\pm$  0.008 \\
NGC~6535 & 15.65 &  0.554 $\pm$  0.041 &  0.127 &  1.457 $\pm$  0.011 &  0.903 $\pm$  0.042 \\
NGC~6541 & 15.15 &  0.112 $\pm$  0.010 &  0.107 &  1.076 $\pm$  0.004 &  0.964 $\pm$  0.011 \\
NGC~6584 & 16.40 &  0.640 $\pm$  0.061 &  0.215 &  1.048 $\pm$  0.003 &  0.408 $\pm$  0.062 \\
NGC~6624 & 15.85 &  1.248 $\pm$  0.003 &  0.025 &  1.383 $\pm$  0.012 &  0.135 $\pm$  0.012 \\
NGC~6637 & 15.75 &  1.092 $\pm$  0.002 &  0.020 &  1.230 $\pm$  0.002 &  0.138 $\pm$  0.003 \\
NGC~6652 & 15.77 &  1.030 $\pm$  0.004 &  0.021 &  1.171 $\pm$  0.004 &  0.141 $\pm$  0.006 \\
NGC~6656 & 14.15 &  0.438 $\pm$  0.011 &  0.141 &  1.373 $\pm$  0.004 &  0.935 $\pm$  0.012 \\
NGC~6681 & 15.63 &  0.074 $\pm$  0.010 &  0.112 &  1.060 $\pm$  0.002 &  0.986 $\pm$  0.010 \\
NGC~6717 & 15.53 &  0.268 $\pm$  0.021 &  0.128 &  1.216 $\pm$  0.003 &  0.948 $\pm$  0.021 \\
NGC~6723 & 15.30 &  0.704 $\pm$  0.039 &  0.301 &  1.075 $\pm$  0.003 &  0.371 $\pm$  0.039 \\
NGC~6752 & 13.70 & -0.058 $\pm$  0.009 &  0.102 &  1.024 $\pm$  0.003 &  1.082 $\pm$  0.010 \\
NGC~6779 & 16.15 &  0.305 $\pm$  0.007 &  0.110 &  1.220 $\pm$  0.003 &  0.914 $\pm$  0.008 \\
NGC~6809 & 14.35 &  0.154 $\pm$  0.021 &  0.097 &  1.060 $\pm$  0.004 &  0.906 $\pm$  0.021 \\
NGC~6838 & 14.21 &  1.185 $\pm$  0.009 &  0.021 &  1.291 $\pm$  0.005 &  0.106 $\pm$  0.010 \\
NGC~6934 & 16.78 &  0.355 $\pm$  0.046 &  0.251 &  1.066 $\pm$  0.005 &  0.711 $\pm$  0.046 \\
NGC~6981 & 16.73 &  0.537 $\pm$  0.122 &  0.247 &  1.008 $\pm$  0.007 &  0.471 $\pm$  0.122 \\
NGC~7078 & 15.75 &  0.131 $\pm$  0.028 &  0.196 &  1.007 $\pm$  0.002 &  0.876 $\pm$  0.028 \\
NGC~7089 & 15.95 &  0.043 $\pm$  0.009 &  0.148 &  0.993 $\pm$  0.002 &  0.951 $\pm$  0.009 \\
NGC~7099 & 15.12 &  0.087 $\pm$  0.005 &  0.077 &  0.960 $\pm$  0.004 &  0.872 $\pm$  0.006 \\
Pal~12   & 16.90 &  0.908 $\pm$  0.005 &  0.008 &  1.019 $\pm$  0.016 &  0.111 $\pm$  0.017 \\
Terzan~7 & 17.67 &  1.028 $\pm$  0.004 &  0.012 &  1.115 $\pm$  0.005 &  0.087 $\pm$  0.007 \\
Terzan~8 & 17.90 &  0.230 $\pm$  0.011 &  0.111 &  1.095 $\pm$  0.011 &  0.865 $\pm$  0.015 \\
\cutinhead{Additional GCs}
AM-1     & 20.92 &  0.799 $\pm$  0.008 &  0.021 &  0.991 $\pm$  0.005 &  0.192 $\pm$  0.009 \\
Eridanus & 20.23 &  0.829 $\pm$  0.015 &  0.017 &  1.009 $\pm$  0.033 &  0.180 $\pm$  0.036 \\
NGC~2419 & 20.35 &  0.198 $\pm$  0.008 &  0.155 &  1.017 $\pm$  0.002 &  0.819 $\pm$  0.008 \\
Pal~3    & 20.40 &  0.688 $\pm$  0.021 &  0.087 &  1.002 $\pm$  0.019 &  0.314 $\pm$  0.028 \\
Pal~4    & 20.65 &  0.816 $\pm$  0.008 &  0.018 &  0.999 $\pm$  0.006 &  0.183 $\pm$  0.010 \\
Pal~14   & 20.00 &  0.828 $\pm$  0.021 &  0.035 &  1.056 $\pm$  0.027 &  0.228 $\pm$  0.034 \\
\enddata                                                                 
\tablecomments{MAD =  Mean Absolute Deviation}
\end{deluxetable}


\begin{deluxetable}{lrrrrr}
\tabletypesize{\footnotesize}
\tablecolumns{6}
\tablewidth{0pc}
\tablecaption{Results from Isochrone Fitting\label{age}}
\tablehead{\colhead{Name}&\colhead{$\feh$}&\colhead{$\afe$}&\colhead{$DM_{F814W}$}&\colhead{E(6$-$8)}&\colhead{Age (Gyr)}}
\startdata
Arp~2    & -1.80 & 0.2 & 17.55 & 0.113 & 13.00 $\pm$ 0.75 \\
Lyng\aa7 & -0.60 & 0.2 & 15.80 & 0.713 & 12.50 $\pm$ 1.00 \\
NGC~104  & -0.70 & 0.2 & 13.30 & 0.023 & 12.75 $\pm$ 0.50 \\
NGC~288  & -1.40 & 0.4 & 14.85 & 0.013 & 12.50 $\pm$ 0.50 \\
NGC~362  & -1.30 & 0.2 & 14.80 & 0.023 & 11.50 $\pm$ 0.50 \\
NGC~1261 & -1.35 & 0.2 & 16.10 & 0.013 & 11.50 $\pm$ 0.50 \\
NGC~2298*& -1.90 & 0.2 & 15.43 & 0.237 & 13.00 $\pm$ 1.00 \\
NGC~3201*& -1.50 & 0.2 & 13.90 & 0.268 & 12.00 $\pm$ 0.75 \\
NGC~4147 & -1.70 & 0.2 & 16.48 & 0.018 & 12.75 $\pm$ 0.75 \\
NGC~4590 & -2.30 & 0.2 & 15.30 & 0.056 & 13.00 $\pm$ 1.00 \\
NGC~4833*& -2.30 & 0.4 & 14.84 & 0.353 & 13.00 $\pm$ 1.25 \\
NGC~5024 & -2.00 & 0.2 & 16.43 & 0.023 & 13.25 $\pm$ 0.50 \\
NGC~5053 & -2.40 & 0.2 & 16.32 & 0.021 & 13.50 $\pm$ 0.75 \\
NGC~5272 & -1.60 & 0.2 & 15.08 & 0.018 & 12.50 $\pm$ 0.50 \\
NGC~5286*& -1.70 & 0.0 & 15.75 & 0.263 & 13.00 $\pm$ 1.00 \\
NGC~5466 & -2.10 & 0.2 & 16.12 & 0.023 & 13.00 $\pm$ 0.75 \\
NGC~5904 & -1.30 & 0.2 & 14.38 & 0.033 & 12.25 $\pm$ 0.75 \\
NGC~5927*& -0.50 & 0.2 & 15.30 & 0.393 & 12.25 $\pm$ 0.75 \\
NGC~5986*& -1.60 & 0.2 & 15.73 & 0.295 & 13.25 $\pm$ 1.00 \\
NGC~6093*& -1.70 & 0.2 & 15.55 & 0.213 & 13.50 $\pm$ 1.00 \\
NGC~6101 & -1.80 & 0.2 & 16.03 & 0.113 & 13.00 $\pm$ 1.00 \\
NGC~6121*& -1.20 & 0.4 & 12.20 & 0.423 & 12.50 $\pm$ 0.50 \\
NGC~6144*& -1.80 & 0.2 & 15.40 & 0.448 & 13.50 $\pm$ 1.00 \\
NGC~6171*& -1.00 & 0.2 & 14.45 & 0.418 & 12.75 $\pm$ 0.75 \\
NGC~6205 & -1.60 & 0.2 & 14.47 & 0.019 & 13.00 $\pm$ 0.50 \\
NGC~6218 & -1.30 & 0.2 & 13.90 & 0.191 & 13.25 $\pm$ 0.75 \\
NGC~6254*& -1.55 & 0.2 & 14.15 & 0.261 & 13.00 $\pm$ 1.25 \\
NGC~6304*& -0.50 & 0.2 & 14.87 & 0.473 & 12.75 $\pm$ 0.75 \\
NGC~6341 & -2.40 & 0.2 & 14.80 & 0.031 & 13.25 $\pm$ 1.00 \\
NGC~6352*& -0.80 & 0.4 & 14.10 & 0.253 & 13.00 $\pm$ 0.50 \\
NGC~6362 & -1.10 & 0.4 & 14.55 & 0.070 & 12.50 $\pm$ 0.50 \\
NGC~6366 & -0.70 & 0.2 & 14.00 & 0.718 & 12.00 $\pm$ 0.75 \\
NGC~6397 & -2.10 & 0.2 & 12.40 & 0.183 & 13.50 $\pm$ 0.50 \\
NGC~6496 & -0.50 & 0.2 & 15.35 & 0.213 & 12.00 $\pm$ 0.75 \\
NGC~6535 & -1.90 & 0.2 & 14.85 & 0.443 & 13.25 $\pm$ 1.00 \\
NGC~6541*& -1.90 & 0.4 & 14.68 & 0.118 & 13.25 $\pm$ 1.00 \\
NGC~6584 & -1.40 & 0.2 & 15.85 & 0.078 & 12.25 $\pm$ 0.75 \\
NGC~6624*& -0.50 & 0.0 & 15.05 & 0.253 & 13.00 $\pm$ 0.75 \\
NGC~6637*& -0.70 & 0.2 & 15.05 & 0.163 & 12.50 $\pm$ 0.75 \\
NGC~6652 & -0.75 & 0.2 & 15.05 & 0.113 & 13.25 $\pm$ 0.50 \\
NGC~6681 & -1.50 & 0.2 & 15.05 & 0.098 & 13.00 $\pm$ 0.75 \\
NGC~6717*& -1.10 & 0.2 & 14.78 & 0.203 & 13.00 $\pm$ 0.75 \\
NGC~6723 & -1.00 & 0.2 & 14.67 & 0.073 & 12.75 $\pm$ 0.50 \\
NGC~6752 & -1.50 & 0.2 & 13.26 & 0.053 & 12.50 $\pm$ 0.75 \\
NGC~6779*& -2.20 & 0.2 & 15.65 & 0.248 & 13.50 $\pm$ 1.00 \\
NGC~6809 & -1.80 & 0.2 & 13.88 & 0.113 & 13.50 $\pm$ 1.00 \\
NGC~6838*& -0.70 & 0.2 & 13.40 & 0.223 & 12.50 $\pm$ 0.75 \\
NGC~6934 & -1.55 & 0.2 & 16.23 & 0.108 & 12.00 $\pm$ 0.75 \\
NGC~6981 & -1.50 & 0.2 & 16.20 & 0.048 & 12.75 $\pm$ 0.75 \\
NGC~7078 & -2.40 & 0.2 & 15.40 & 0.083 & 13.25 $\pm$ 1.00 \\
NGC~7089 & -1.60 & 0.2 & 15.48 & 0.048 & 12.50 $\pm$ 0.75 \\
NGC~7099 & -2.40 & 0.2 & 14.82 & 0.053 & 13.25 $\pm$ 1.00 \\
Pal~12   & -0.80 & 0.0 & 16.40 & 0.033 &  9.50 $\pm$ 0.75 \\
Terzan~7 & -0.60 & 0.0 & 17.15 & 0.073 &  8.00 $\pm$ 0.75 \\
Terzan~8 & -2.40 & 0.4 & 17.50 & 0.133 & 13.50 $\pm$ 0.50 \\
\cutinhead{Additional GCs [$DM_V$ and E($V-I$)]}
AM-1     & -1.50 & 0.2 & 20.41 & 0.020 & 11.10 $\pm$ 0.50 \\
Eridanus & -1.30 & 0.2 & 19.78 & 0.060 & 10.50 $\pm$ 0.50 \\
NGC~2419 & -2.00 & 0.2 & 20.05 & 0.062 & 13.00 $\pm$ 1.00 \\
Pal~3    & -1.50 & 0.2 & 19.89 & 0.060 & 11.30 $\pm$ 0.50 \\
Pal~4    & -1.30 & 0.2 & 20.14 & 0.055 & 10.90 $\pm$ 0.50 \\
Pal~14   & -1.50 & 0.2 & 19.51 & 0.045 & 10.50 $\pm$ 0.50 \\
\enddata
\tablecomments{An asterisk (*) after the name indicates that the differential reddening
corrected CMD was used in the isochrone analysis. $E(6-8)$ = $E(F606W-F814W)$.}
\end{deluxetable}

\clearpage

\begin{deluxetable}{ccc}
\tablecolumns{3}
\tablewidth{0pc}
\tablecaption{Fitting Function Coefficients\label{coeff}}
\tablehead{\colhead{Name}&\colhead{This Paper}&\colhead{Paper VII}}
\startdata
$a_0$&$ 0.947\pm0.016$ &$ 0.946\pm0.015$ \\
$a_1$&$ 0.809\pm0.017$ &$ 0.809\pm0.017$ \\
$a_2$&$ 0.900\pm0.021$ &$ 1.012\pm0.036$ \\
$a_3$&$ 0.022\pm0.011$ &$ 0.029\pm0.019$ \\
$b_0$&$-0.098\pm0.023$ &$-0.091\pm0.020$ \\
$b_1$&$-0.244\pm0.050$ &$-0.253\pm0.051$ \\
$b_2$&$-0.105\pm0.020$ &$-0.127\pm0.025$ \\
\enddata
\end{deluxetable}

\end{document}